\documentclass[traditabstract]{aa}

\usepackage{natbib}
\usepackage{graphicx}
\usepackage{lscape}
\usepackage{times}

\newcommand\swift{{\it Swift}} 
 
\newcommand\invsqrcm{cm$^{-2}$} 
\newcommand\nh{$N_H$}
\newcommand\nhx{$N_{H,x}$}
\newcommand\av{$A_V$}

\newcommand\rv{$R_V$}

\newcommand\MWbump{2175\AA}
\def\lesssim{\mathrel{\hbox{\rlap{\hbox{\lower4pt\hbox{$\sim$}}}\hbox{$<$}}}}
\def\gtrsim{\mathrel{\hbox{\rlap{\hbox{\lower4pt\hbox{$\sim$}}}\hbox{$>$}}}}

\begin{document}
 
\title{The Dust Extinction Curves of Gamma-Ray Burst Host Galaxies}

\author{P.~Schady\inst{1,2}, T.~Dwelly\inst{3}, M.J.~Page\inst{2}, T.~Kr{\"u}hler\inst{1,4,5}, J.~Greiner\inst{1}, S.R.~Oates\inst{2}, M.~De~Pasquale\inst{2}, M.~Nardini\inst{1,6}, P.W.A.~Roming\inst{7}, A.~Rossi\inst{8} and M.~Still\inst{9}}
\institute{Max-Planck Institut f{\" u}r Extraterrestrische Physik, Giessenbachstra\ss e 1, 85748 Garching, Germany
\and
The UCL Mullard Space Science Laboratory, Holmbury St Mary, Dorking, Surrey, RH5 6NT, UK
\and
Physics and Astronomy, University of Southampton, Highfield, Southampton, SO17 1BJ, UK
\and
Universe Cluster, Technische Universit\"{a}t M\"{u}nchen, Boltzmannstra{\ss}e 2, 85748, Garching, Germany
\and
Dark Cosmology Centre, Niels Bohr Institute, University of Copenhagen, Juliane Maries Vej 30, 2100 Copenhagen, Denmark
\and
Deparment of Physics, University of Milano-Bicocca, IT 20126 Milano, Italy
\and
Space Science \& Engineering Division, Southwest Research Institute, 6220 Culebra Rd, San Antonio, TX 78238, USA
\and
Th{\"u}ringer Landessternwarte Tautenburg, Sternwarte 5, 07778 Tautenburg, Germany
\and
NASA Ames Research Center, M/S 244-30, Moffett Field, CA 94035, USA
}

\date{Accepted: Received: }

\abstract
{The composition and amount of interstellar dust within gamma-ray burst (GRB) host galaxies is of key importance when addressing selection effects in the GRB redshift distribution, and when studying the properties of their host galaxies. As well as the implications for GRB research, probing the dust within the high-z hosts of GRBs also contributes to our understanding of the conditions of the interstellar medium and star-formation in the distant Universe. Nevertheless, the physical properties of dust within GRB host galaxies continues to be a highly contended issue. In this paper we explore the mean extinction properties of dust within the host galaxies of a sample of 17 GRBs with total host galaxy visual extinction $A_V<1$ ($\langle A_V\rangle$=0.4), covering a redshift range $z=0.7-3.1$. We find the average host extinction curve to have an ultraviolet slope comparable to that of the LMC, but with little evidence of a \MWbump\ dust extinction feature as observed along Milky Way and LMC sightlines. We cannot at present rule out the presence of a \MWbump\ feature, and both the standard SMC and LMC extinction curves also provide good fits to our data. However, we {\em can} reject an extinction curve that has a UV slope as flat as the mean Milky Way extinction curve, whilst also having a \MWbump\ feature as prominent as seen in the mean Milky Way extinction curve. This is in contrast to the clear detection of a \MWbump\ bump and the flatter extinction curves of some more heavily extinguished GRBs ($A_V>1$), which may be indicative of there being a dependence between dust abundance and the wavelength dependence of dust extinction, as has been previously speculated.
\keywords{gamma-rays: bursts - gamma-ray: observations - galaxies: ISM - dust, extinction}}

\titlerunning{The Dust Extinction Curve of Gamma-Ray Burst Hosts}
\authorrunning{P.~Schady et al.}
\maketitle

\section{Introduction}
\label{sec:intro}
The effects of interstellar dust through its extinction and emission properties permeates into most aspects of astrophysics, either as an indirect measure of the environmental conditions and enrichment processes present within a galaxy, or as a hindrance to be removed from observations.

The dust extinction along numerous lines-of-sight within the Milky Way \citep[e.g.][]{sm79} and the Small and Large Magellanic Clouds \citep[SMC and LMC, respectively;][]{fit89} have been well studied, both in terms of the absolute extinction, $A_\lambda$, at a given wavelength $\lambda$, and in terms of the relative $A_{\rm B}-A_{\rm V}$ extinction, or reddening, $E(B-V)$. The bulk of this work has been focused on determining the relative extinction as a function of wavelength, or the {\em extinction law} or {\em curve} \citep{ccm89,pei92,gcm+03,fit04,cla04}, which holds information on the grain size distribution and composition of the extinguishing dust along the observed line-of-sight. However, it has thus far only been possible to measure the dust extinction curve in extragalactic environments along lines-of-sight to bright background quasars \citep[e.g.][]{wfp+95,pfb+91,ml04,mso+04,ehl+05,vcl+06,gmj+10} or gamma-ray bursts \citep[GRBs; e.g.][]{gw01,sff03,sfa+04,kkz06,sww+07,wfl+06,smp+07}.

GRBs have been particularly effective at probing the extinction curve of high-z galaxies thanks to their intrinsically simple spectra \citep{clw06,hlp+08,llw08,pbb+08,pbk+10,pmu+11,zwf+11}. Other efforts to measure the dust extinction properties in extragalactic environments have been made from stacked spectra of large samples of galaxies of similar type \citep[e.g.][]{cks94,vwg03,np05,npp+07,npc+09}. However, such observations are subject to complex radiative transfer effects that reprocess the stellar light through numerous episodes of gas and dust absorption, emission and scattering. As such, the derived galaxy dust extinction properties are highly dependent on the geometric distribution of the dust, gas and stars assumed. The measured wavelength dependent stellar attenuation is then referred to as the galaxy's attenuation curve \citep[e.g. Calzetti law;][]{cks94}, which is distinct from its dust extinction curve \citep[e.g. CCM extinction law;][]{ccm89}.

A great deal of progress has been made in the theoretical modelling of the extinction law of dust produced by core collapse supernovae (SNe) \citep[e.g.][]{tf01,nku+03,sfs04,bs07}, which are thought to have been a possible dominant source of dust formation at high redshift ($z > 6$), when the contribution to dust formation from asymptotic giant branch stars was still small. Indeed, \citet{gmj+10} found evidence for differences in the dust extinction curves in high redshift ($z>4$) quasars relative to lower redshift quasar extinction curves, suggestive of a change in the dust formation or processing mechanism at these high redshifts. SN dust extinction curves have also been claimed to have been measured in the hosts of high-z quasars \citep[SDSSJ104845.05+463718.3, z=$6.193$;][]{mso+04} and GRBs (GRB~050904, Stratta et al. 2007, although see Zafar et al. 2010; GRB~071025, Perley et al. 2010).

A further ongoing and poorly understood issue is the origin and prevalence of the dust extinction feature observed at \MWbump\ in the Milky Way extinction law, and at lower significance also in the Large Magellanic Cloud (LMC). Small carbonaceous dust grains are thought to play a role \citep{dl84}, although it is still not clear what can produce the observed variation in width and height of this feature without varying its central wavelength, which shows very small scatter. Observationally, a failure to account for this feature can also result in a significant under-estimation of the amount of dust extinction in some sources \citep[e.g.][]{csb10}. A more detailed investigation of the variation in dust extinction properties with redshift and with environmental conditions is therefore important to identify the carrier of this broad extinction bump, which may have implications for dust formation.

The vast luminosity, simple intrinsic spectra and fading nature of GRBs make them unique and highly effective probes of the extinction properties of dust in high-z galaxies. This is particularly the case for long GRBs \citep[$\gtrsim 2.0$~sec,][]{kmf+93}, which are associated with massive star formation (e.g. GRB~980425; Galama et al. 1999, GRB~030329; Hjorth et al. 2003, Stanek et al. 2003), and thus provide lines-of-sight to star-forming regions within the host galaxy. Short GRBs typically occur far from the centre of their host galaxies, and their faint afterglows provide a less effective probe of their (probably low density) surrounding environment. In this paper we therefore focus on long GRBs only, and all references to GRBs beyond this point correspond to this class of burst. 

In this current era of rapid-response GRB observations, the number of well-monitored GRBs are now becoming sufficiently large to allow statistically meaningful analysis to be done on the characteristic properties of GRBs and their host galaxies. Early-time, multi-wavelength observations taken with the X-ray Telescope \citep[XRT;][]{bhn+05} and the Ultraviolet and Optical Telescope \citep[UVOT;][]{rkm+05} onboard \swift\ \citep{gcg+04}, in addition to optical and near-infrared (NIR) data from rapid-response ground based telescopes, provide broadband, high signal-to-noise spectral energy distributions (SEDs) for large numbers of afterglows. Dust extinction is typically larger at bluer wavelengths, and GRB afterglows are, therefore, increasingly attenuated at optical and ultraviolet (UV) wavelengths relative to their NIR afterglow. The relatively unattenuated rest-frame NIR and X-ray bands ($> 2$~keV) therefore allow the intrinsic power-law or broken power-law continua to be well-determined, from which the imprint left by dust extinction can be measured to a high level of accuracy. The blue UVOT filters provide unique coverage of the dust-extinction profile on GRB afterglows at $z<1.5$. At higher redshifts, redder wavelength observations from ground-based instruments such as the Gamma-Ray Burst Optical and Near-Infrared Detector \citep[GROND;][]{gbc+08} become important. This longer wavelength coverage is also critical for measuring heavily extinguished GRBs which may not be detected by UVOT.

Typically it is found that for GRBs with well-observed afterglow SEDs, the amount of host galaxy extinction is small (total $V$-band or {\em visual} extinction $\langle A_V\rangle<0.3$) \citep[e.g.][]{kkz06,spo+10, sww+07,gkk+11}, and has a NIR to X-ray wavelength dependence similar to that of the featureless mean Small Magellanic Cloud (SMC). However, recent dedicated GRB rapid-response programmes with NIR wavelength coverage are now revealing a sample of GRBs ($\sim 10$\%) that have been significantly extinguished ($A_V>1$) by host galaxy dust \citep{gkk+11}, many of which show evidence for a much flatter, Milky Way-like host galaxy extinction curve than is observed in the hosts of only moderately extinguished GRBs. There are now four GRBs with spectroscopically confirmed detections of the \MWbump\ absorption bump at the redshift of the GRB host galaxy (GRB~070802; El{\'i}asd{\'o}ttir et al. 2009, GRB~080607; Prochaska et al. 2009, GRB~080605 and 080805; Zafar et al. 2011), as well as several GRBs with a \MWbump\ extinction feature detected in the GRB SED (GRB~050802; Oates et al. 2007, GRB~070802; Kr{\"u}hler et al. 2008, GRB~080603A; Guidorzi et al. 2011). Several claims have also been made for GRB hosts having a `grey' extinction law, where the dust extinction is only weakly dependent on wavelength \citep[e.g.][]{sff03,spl+05,pbb+08}. The causes for this range in the total host galaxy dust extinction, and in the shape of the extinction curves of GRB hosts are still poorly understood, and better comprehension on this issue requires the systematic analysis of a homogenous data set of GRB afterglows.

In this paper we explore the variation in the dust extinction properties of star-forming galaxies by performing a simultaneous analysis of a sample of 17 NIR to X-ray GRB afterglow SEDs. We wish to study the wavelength dependence of the extinction curve within GRB host galaxies rather than the absolute amount of GRB host galaxy reddening, and we are thus interested in analysing the afterglow SEDs over a broad bandpass. Our sample has a large redshift range ($z=0.7-3.1$), which provides coverage of the rest-frame wavelength dependence of extinction over several decades. As well as addressing the prominence of a \MWbump\ feature and the presence of grey dust in GRB host galaxies, the analysis presented in this paper also investigates the uniformity of GRB host extinction laws in general. The existence of an extinction law specific to GRB host galaxies would not only improve GRB afterglow SED modelling and thus the accuracy of host galaxy visual extinction measurements, but it would also hold information on the grain properties of the dust in the GRB surrounding environment.

In section~\ref{sec:red} we give our sample selection criteria and describe our data reduction and afterglow SED analysis. Based on our individual afterglow SED analysis, we select an optimal sample of GRBs with which to study the dust extinction properties of GRB host galaxies. This final sample and the consequent simultaneous SED analysis is described in section~\ref{sec:method}. Our results are presented in section~\ref{sec:results}, and a discussion and the main conclusions from our work are provided in sections~\ref{sec:disc} and \ref{sec:conc}. Throughout the paper temporal and spectral indices, $\alpha$ and $\beta$, respectively, are denoted such that $F(\nu,t)\propto \nu^{-\beta}t^{-\alpha}$, and all errors are $1\sigma$ unless specified otherwise.

\begin{tiny}
\begin{table*}
\setcounter{table}{0}
\caption{Table listing the 49 GRBs chosen by our selection criteria to have well-sampled afterglow SEDs.\label{tab:GRBsamp}}
\begin{center}
\begin{tabular}{llccllll}
\hline
GRB & z & $N_{H,Gal}$ &  $A_{V,Gal}$ & Epoch$^\dag$ & UV/optical/NIR bandpasses & Rest-frame Band & Best-fit Model \\
 & & ($10^{21}$~\invsqrcm) & (mag) & (ks) & & Coverage (\AA) & \\
\hline\hline
050319 & 3.24$^a$ & 0.13 & 0.03 & 7 & $I^{1,2}R^{1-3}v~b$ & 920--2090 & bknp/MW \\
050525A & 0.606$^b$ & 0.91 & 0.29 & 20 & $K^{4,5}H^4J^{4,6}I^{4,7}R^{4,8,9}v~b~u~w1~m2~w2$ & 1000-14540 & pow/SMC \\
050730 & 3.969$^c$ & 0.30 & 0.16 & 10 & $(KJIr)^{10}v~b$ & 790--4700 & pow/MW \\
050802 & 1.71$^d$ & 0.19 & 0.06 & 2 & $v~b~u~w1~m2~w2$ & 590--2160 & pow/MW \\
050820A & 2.615$^e$ & 0.44 & 0.14& 10 & $J^{11}(zIRg)^{12}v~b~u~w1$ & 620--3710 & bknp/LMC \\
050922C & 2.198$^f$ & 0.54  & 0.32 & 5 & $R^{13-18}v~b~u~w1$ & 710--2360 & pow/MW \\
060206 & 4.048$^g$ & 0.09 & 0.04 & 20 & $(KHJ)^{19}R^{20-22}v~b$ & 770--4630 & bknp/MW \\
060418 & 1.49$^h$ & 0.88 & 0.69 & 5 & $(KHJ)^{23}z^{23,24}I^{25}R^{26}v~b~u~w1~m2~w2$ & 640--9380 & bknp/SMC \\
060502A & 1.51$^i$ & 0.35 & 0.10 & 5 & $R^{27}v~b~u~w1$ & 900--3010 & pow/SMC \\
060512 & 2.1$^j$ & 0.15 & 0.05 & 5 & $Ks^{28}J^{29}R^{30,31}v~b~u$ & 1240--7530 & pow/SMC \\
060526 & 3.21$^k$ & 0.50 & 0.21 & 6.5 & $I^{32,33}R^{33-40}v~b$ & 930--2100 & pow/MW \\
060607A & 3.082$^l$ & 0.24 & 0.09 & 8 & $H^{23}(irg)^{41}v~b~u$ & 950--4200 & pow/SMC Ä\\
060729 & 0.54$^m$ & 0.45 & 0.17 & 70 & $i^{m}R^{42}v~b~u~w1~m2~w2$ & 1040--5650 & bknp/SMC \\
060904B & 0.703$^n$ & 1.13 & 0.53 & 5 & $(KJI)^{43}R^{44}v~b~u~w1~m2~w2$ & 940--13710 & pow/LMC \\
060908 & 2.43$^o$ & 0.23 & 0.09 & 6 & $R^v~b~u~w1$ & 660--2200 & pow/SMC \\
060912 & 0.937$^p$ & 0.39 & 0.16& 1 & $v~b~u~w1~w2$ & 830--3020 & pow/MW \\
061007 & 1.262$^q$ & 0.18 & 0.06 & 0.7 & $(iR)^{47}v~b~u~w1~m2~w2$ & 710--3850 & pow/SMC \\
061121 & 1.314$^r$ & 0.40 & 0.14 & 6 & $I^{48-50}R^{51}v~b~u~w1~m2~w2$ & 690--3830 & bknp/LMC \\
061126 & 1.159$^s$ & 1.02 & 0.56 & 2 & $i^{52}R^{s,53,54}v~b~u~w1~m2$ & 740--4030 & bknp/LMC \\
070110 & 2.352$^t$ & 0.16  & 0.04 & 10 & $R^{55}v~b~u$ & 1150--2250 & bknp/SMC \\
070318 & 0.836$^u$ & 0.14 & 0.05 & 6 & $v~b~u~w1~m2~w2$ & 870--3190 & pow/SMC \\
070810A & $2.17^v$ & 0.18 & 0.07 & 5 & $R^{v}v~b~u$ & 1220--2380 & pow/MW \\
071112C & $0.823^w$ & 0.74 & 0.36 & 1.3 & $(KJ)^{56}I^{57}R^{57-59}v~g^{57}b~u~w1~m2~w2$ & 880--12810 & pow/SMC \\
080210 & $2.641^x$ & 0.55 & 0.26 & 5 & $KHJz'i'r'g'$ & 1060--6370 & bknp/LMC \\
080319B & $0.937^y$ & 0.11 & 0.03 & 200 & $(ir)^{60}b~u~w1~m2~w2$ & 830--4490 & pow/SMC \\
\hline
\end{tabular}
\end{center}
NOTE- Columns list GRB redshift, Galactic column density, $N_{H,Gal}$, and visual extinction, $A_{V,Gal}$, in the line-of-sight to the GRB, the corresponding SED epoch, the UV/optical/NIR band passes included in the GRB afterglow SED, the rest-frame coverage of the SED, and the best-fit SED model. In this last column we indicate both the best-fit continuum (power-law or broken power-law abbreviated to pow and bknp respectively) and host galaxy dust extinction model, where we have used the abbreviation MW to indicate Milky Way extinction curve model.\\
 $^a$ \citet{fhj+05};
 $^b$ \citet{fcb+05};
 $^c$ \citet{sve+05};
 $^d$ \citet{fsj+05};
 $^e$ \citet{lve+05};
 $^f$ \citet{jfp+05};
 $^g$ \citet{pwf+06};
 $^h$ \citet{dfp+06}
 $^i$ \citet{cpf+06};
 $^j$ \citet{stf+06};
 $^k$ \citet{bg06};
 $^l$ \citet{lvs+06};
 $^m$ \citet{tlj+06};
 $^n$ \citet{fdm+06};
 $^o$ \citet{rjt+06};
 $^p$ \citet{jlc+06};
 $^q$ \citet{jft+06};
 $^r$ \citet{bpc06};
 $^s$ \citet{pbb+08};
 $^t$ \citet{jmf+07};
 $^u$ \citet{jfa+07};
 $^v$ \citet{tpc+07};
 $^w$ \citet{jfv+07};
 $^x$ \citet{jvm+08};
 $^y$ \citet{vsm+08};\\
 $^\dag$ relative to time that BAT triggered on the GRB\\
$^1$ \citet{huk+07};
$^2$ \citet{krm07};
$^3$ \citet{wvw+05};
$^4$ \citet{cb05};
$^5$ \citet{rg05};
$^6$ \citet{fhs+05}; 
$^7$ \citet{ytk05};
$^8$ \citet{hhg+05};
$^{9}$ \citet{mbs05},
$^{10}$ \citet{pcm+06};
$^{11}$ \citet{meb+05};
$^{12}$ \citet{ckh+06};
$^{13}$ \citet{dp05}; 
$^{14}$ \citet{jpt+05};
$^{15}$ \citet{ap05};
$^{16}$ \citet{hkh+05};
$^{17}$ \citet{pmm+05};
$^{18}$ \citet{dpf+05}; 
$^{19}$ \citet{apb06};
$^{20}$ \citet{cvw+07};
$^{21}$ \citet{sdp+07};
$^{22}$ \citet{wvw+06};
$^{23}$ \citet{mvm+07}; 
$^{24}$ \citet{nif+06}; 
$^{25}$ \citet{cob06a}; 
$^{26}$ \citet{kop06};
$^{27}$ \citet{cof06}, 
$^{28}$ \citet{hlm+06}; 
$^{29}$ \citet{sdp+06}; 
$^{30}$ \citet{cen06a};
$^{31}$ \citet{mil06};
$^{32}$ \citet{cob06b}; 
$^{33}$ \citet{tgb+06}; 
$^{34}$ \citet{bgv+06};
$^{35}$ \citet{cig+06},
$^{36}$ \citet{dhm+07}; 
$^{37}$ \citet{gtb+06};
$^{38}$ \citet{kbs+06};
$^{39}$ \citet{md06};
$^{40}$ \citet{rpi+06}
$^{41}$ \citet{nrc+09}; 
$^{42}$ \citet{qr06}; 
$^{43}$ \citet{cb06};
$^{44}$ \citet{skv06}
$^{45}$ \citet{act+06};
$^{46}$ \citet{wtr06};
$^{47}$ \citet{mmg+07}; 
$^{48}$ \citet{cen06b};
$^{49}$ \citet{cob06c};
$^{50}$ \citet{tor06};
$^{51}$ \citet{uau06};
$^{52}$ \citet{gkg+08};
$^{53}$ \citet{smg+06};
$^{54}$ \citet{wm06};
$^{55}$ \citet{mjv07};
$^{56}$ \citet{usa+07};
$^{57}$ \citet{yys+07};
$^{58}$ \citet{dmm+07};
$^{59}$ \citet{bkp+07};
$^{60}$ \citet{rks+08};\\
\end{table*}
\end{tiny}

\begin{tiny}
\begin{table*}
\setcounter{table}{0}
\caption{Continued}
\begin{center}
\begin{tabular}{llccllll}
\hline
GRB & z & $N_{H,Gal}$ &  $A_{V,Gal}$ & Epoch & UV/optical/NIR bandpasses & Rest-frame Band & Best-fit Model \\
 & & ($10^{21}$~\invsqrcm) & (mag) & (ks) & & Coverage (\AA) & \\
\hline\hline
080319C & $1.95^a$ & 0.22 & 0.08 & 0.5 & $R^1v~b~u$ & 1310--2260 & pow/SMC \\
080411 & $1.03^b$ & 0.58 & 0.10 & 8 & $KHJz'i'r'g'$ & 1900--11420 & bknp/MW \\
080413B & $1.10^c$ & 0.31 & 0.11 & 0.65 & $v~b~u~w1~m2~w2$ & 760--2790 & bknp/MW \\
080430 & $0.767^d$ & 0.10 & 0.04 & 10 & $I^{2,3}R^{3-5}v~b~u~w1~m2~w2$ & 910--5010 & bknp/SMC \\
080603B & $2.69^e$ & 0.12 & 0.04 & 6.5 & $(KHJ)^6R^{7,8}b~u$ & 1050--6330 & pow/SMC \\
080710 & $0.845^f$ & 0.41 & 0.23 & 10 & $(KHJz'i'r')^9v~g'^9b~u~w1~m2~w2$ & 870--12560 & pow/SMC \\
080721 & $2.591^g$ & 0.69 & 0.31 & 6 & $R^{10}v~b~u$ & 1070--2100 & bknp/MW \\
080804 & $2.204^h$ & 0.16 & 0.05 & 1.5 & $KHJz'i'r'v~g'b~u$ & 1200--7230 & bknp/MW \\
080810 & $3.35^i$ & 0.33 & 0.09 & 5 & $I^{11}R^{11,12}v~g^{11}b$ & 900--2030 & pow/SMC \\
080916A & $0.689^j$ & 0.18 & 0.06 & 6 & $v~b~u~w1~m2~w2$ & 950--3470 & bknp/SMC \\
080928 & $1.692^k$ & 0.56 & 0.21 & 40 & $b~u~w1~m2~w2$ & 590--1810 & pow/SMC \\
081008 & $1.968^l$ & 0.71 & 0.29 & 1 & $v~b~u~w1$ & 760--1970 & pow/LMC \\
081121 & $2.512^m$ & 0.40 & 0.16 & 10 & $KHJz'i'r'v~g'b~u$ & 1100--6600 & bknp/SMC \\
081203A & $2.05^n$ & 0.17 & 0.06 & 10 & $I^{13,14}R^{13-16}v~g^{15,16}b~u~w1$ & 740--2900 & bknp/MW \\
081222 & $2.77^o$ & 0.22 & 0.06 & 0.7 & $KHJz'i'r'v~g'b~u~w1$ & 600--6150 & bknp/MW \\
090102 & $1.546^p$ & 0.41 & 0.15 & 1 & $KHJz'i'r'v~g'b$ & 1530--9100 & bknp/MW \\
090424 & $0.544^q$ & 0.19 & 0.08 & 1 & $(IR)^{17}v~b~u~w1~m2~w2$ & 1040--5730 & bknp/MW \\
090618 & $0.54^r$ & 0.58 & 0.27 & 2 & $v~b~u~w1~m2~w2$ & 1040--3800 & bknp/SMC \\
090927 & $1.37^s$ & 0.29 & 0.10 & 9 & $v~b~u~w1~m2~w2$ & 680--2470 & pow/MW \\
091018 & $0.971^t$ & 0.28 & 0.09 & 30 & $KHJz'i'r'v~g'b~u~w1~m2~w2$ & 810--11760 & bknp/SMC \\
091020 & $1.71^u$ & 0.14 & 0.05 & 6 & $v~b~u~w1~m2~w2$ & 590--2160 & pow/LMC \\
091029 & $2.752^v$ & 0.11 & 0.05 & 1 & $KHJz'i'r'v~g'b~u~w1$ & 600--6180 & bknp/LMC \\
091127 & $0.490^w$ & 0.28 & 0.12 & 3.7 & $HJz'i'r'w1~m2$ & 1340--11950 & bknp/LMC \\
091208B & $1.063^x$ & 0.49 & 0.16 & 6 & $v~b~u~w1~m2~w2$ & 780--2840 & pow/LMC \\
\hline
\end{tabular}
\end{center}
 $^a$ \citet{wtv+08};
 $^b$ \citet{tcm+08};
 $^c$ \citet{vtm+08};
 $^d$ \citet{cf08a};
 $^e$ \citet{cf08b};
 $^f$ \citet{pcb08};
 $^g$ \citet{jmv+08};
 $^h$ \citet{tuv+08};
 $^i$ \citet{pph+08};
 $^j$ \citet{fmh+08};
 $^k$ \citet{vmf+08};
 $^l$ \citet{cfc+08a};
 $^m$ \citet{br08};
 $^n$ \citet{klp+09};
 $^o$ \citet{cfc+08b};
 $^p$ \citet{ujm+09};
 $^q$ \citet{cpc+09};
 $^r$ \citet{cpj+09};
 $^s$ \citet{lfh+09};
 $^t$ \citet{ugt+09};
 $^u$ \citet{xft+09};
 $^v$ \citet{cpc09};
 $^w$ \citet{cfl+09};
 $^x$ \citet{wtc+09}\\
 $^1$ \citet{lcf08};
 $^2$ \citet{tfu+08};
 $^3$ \citet{tmg+08};
 $^4$ \citet{asb+08};
 $^5$ \citet{oh08};
 $^6$ \citet{mbp08};
 $^7$ \citet{kba08a};
 $^8$ \citet{kba08b};
 $^9$ \citet{kga+09}
 $^{10}$ \citet{srh+09};
 $^{11}$ \citet{yyk+08};
 $^{12}$ \citet{ofn+08};
 $^{13}$ \citet{raa08};
 $^{14}$ \citet{ik08};
 $^{15}$ \citet{vol08};
 $^{16}$ \citet{mns+08};
 $^{17}$ \citet{kkz+10};
\end{table*}
\end{tiny}

\section{Data Sample and Reduction}
\label{sec:red}
\subsection{Sample Selection}
\label{ssec:sample}
To study the extinction curve in the local environment of GRBs, we require a sample of bursts with afterglow data over a broad bandpass, and in particular, covering rest-frame wavelengths $\lesssim2000$\AA, where the wavelength dependence of extinction can vary significantly between environments (e.g. SMC, LMC and Milky Way extinction curves). To carry out our analysis, we select those GRBs from the \swift\ sample up to the end of 2009 that $i)$ are classified as long, $ii)$ have a spectroscopic redshift, $iii)$ were observed with the XRT and the UVOT within one hour of triggering the \swift\ Burst Alert Telescope \citep[BAT;][]{bbc+05}, $iv)$ had an afterglow detected by both XRT and UVOT with a peak UVOT $v$-band magnitude $v<19$, and finally $v)$ were detected in a total of at least four UV/optical/NIR filters, either ground based or UVOT, where at least one of these filters is at a rest-frame wavelength $\lambda<2000$\AA. A spectroscopic redshift is required in order to be able to model the intrinsic GRB afterglow spectrum, and we specify the latter three criteria in order to select GRBs that have sufficiently high quality broadband afterglow SEDs to constrain our analysis. The X-ray data provides a constraint to the afterglow intrinsic broadband spectral shape, and a further four UV-NIR datapoints are required to measure the optical spectral index, $\beta$, the flux normalisation, the total host galaxy visual, or $V$-band ($\lambda\sim 5500$\AA) extinction, \av, along the line-of-sight, and the UV slope of the extinction curve. This selection criteria leaves a sample of 49 GRBs out of a total of 488 GRBs observed by \swift\ by the end of 2009, and they are listed in Table~\ref{tab:GRBsamp}.

By selection, this sample of GRBs will have well-sampled afterglow SEDs from which the intrinsic afterglow spectrum can be accurately modelled. However, to study the extinction curve of their host galaxies, two further selection criteria are required. Firstly, there must be evidence for host galaxy dust extinction, and secondly, the GRB afterglow should have an intrinsic power-law spectrum from the NIR through to X-ray to avoid degeneracy between the afterglow intrinsic spectral slope and the amount of dust extinction. This second level of filtering requires that we first model the SEDs of each of the initial 49 GRBs selected above.

We describe the data reduction and SED analysis on the preliminary sample of 49 GRBs in sections~\ref{ssec:opt}-\ref{ssec:SEDs}. Based on this analysis, in section~\ref{sec:method} we present our final sample of host galaxy dust extinguished GRBs as well as the SED analysis carried out to study the average properties of GRB host galaxy extinction curves.

\subsection{UVOT, GROND, and ground based data}
\label{ssec:opt}
The UVOT contains three optical and three UV filters, which cover the wavelength range between 160~nm and 600~nm, and a clear white filter that covers the wavelength range between 160~nm and 800~nm \citep{pbp+08}. Photometry was carried out on pipeline processed sky images downloaded from the \swift\ data centre\footnotemark[1], 
\footnotetext[1]{http://www.swift.ac.uk/swift\_portal/}
following the standard UVOT procedure \citep{pbp+08}. We applied the tool {\sc uvot2pha} (v1.6) to convert UVOT photometry into spectral files compatible with the spectral fitting package {\sc xspec} \citep{arn96}, and used version v104 of the \swift/UVOT response matrices. In using the UVOT filter response curves, we take the known extended tail, or red leak, in the $uvw1$ ($\lambda_c =2600$) and $uvw2$ ($\lambda_v = 1928$) filters \citep[e.g.][]{brm+10} directly into account.

For 10 of the 49 GRBs in our sample, data at a similar epoch to the \swift\ observations were available from GROND, a seven channel imager covering the 400--2310~nm wavelength range simultaneously ($g'$, $r'$, $i'$, $z'$, $J$, $H$, $K_s$) mounted at the 2.2~m MPI/ESO telescope at the ESO/La Silla observatory in Chile. For these 10 GRBs, the GROND data were reduced using the GROND pipeline \citep{kkg+08}, and light curves were constructed using standard IRAF tasks \citep{tod93} similar to the procedure described in \citet{kkj+08}. GROND optical and NIR photometry was calibrated using SDSS \citep{aaa+09} and 2MASS \citep{scs+06} field stars, respectively. For those GRBs where GROND data were not available, optical or NIR photometry redward of the UVOT bandpass and reported in refereed papers and GCNs were included in the SED ($\sim 50$\% of the sample). For all our data, the calibration systemic error was added in quadrature to the photometric errors, and where no systematic error was available, an error of 0.3~mag was assumed.

Spectral files were produced for each filter using the appropriate responsivity curves. Cousins $R$ and $I$ responsivity curves were taken from \citet{bes90}, and the $J$, $H$ and $K$-band responsivity curves were taken from \citet{cwb+92}, \citet{cww92} and \citet{bcp98}, respectively. In constructing these optical response matrices we have not taken into account the detector quantum efficiency (QE) curves nor atmospheric effects, which we expect to have a small impact on our very broadband spectral modelling. For any non-GROND observations taken in the SDSS $u'g'r'i'z'$ filter-set we have assumed that the response matricies can be well approximated by the SDSS-DR6 system response curves\footnotemark[2], which do included the detector QEs.
\footnotetext[2]{http://www.sdss.org/dr6/instruments/imager/}
We expect that the effects of ignoring the small differences between the nominal SDSS system and the various true filter+CCD QE+atmospheric extinction curves will be very small. GROND responsivity curves were taken from the GROND home page\footnotemark[3] and include the QE of the detectors.
\footnotetext[3]{http://www.mpe.mpg.de/~jcg/GROND/}

\subsection{XRT data}
\label{ssec:xray}
The XRT data were reduced with the \swift\ {\sc xrtpipeline} tool (v0.12.4), and X-ray light curves and spectra in the 0.3--10~keV energy range were extracted from the event data using the software package {\sc xselect} (v2.4). Effective area files were created with the {\sc xrtmkarf} tool (v0.5.6), where exposure maps were used to correct for bad columns, and response matrices were taken from version v011 of the XRT calibration files for both WT and PC mode data. The spectral files were grouped to $\geq 20$ counts per energy channel.

\subsection{SED analysis}
\label{ssec:SEDs}
We created an X-ray through to optical/NIR instantaneous SED for all 49 selected GRBs following the method described in \citet{spo+10}. The epoch of the instantaneous SED was selected to be at a time free of spectral evolution, and was also chosen to minimise the interpolation in time of the optical and X-ray data. Periods of spectral evolution were identified in the X-ray data as those where the light curve hardness ratio\footnotemark[4] deviated significantly (90\% confidence limit) from the mean. 
\footnotetext[4]{http://www.swift.ac.uk/xrt\_curves/}
The UV/optical/NIR data were screened for spectral evolution by verifying the consistency (at 90\% confidence) of the best-fit temporal decay index between filters.

We fitted all GRB afterglow SEDs within {\sc xspec} (v12.5.1), first modelling the intrinsic continuum with a power-law and then with a broken power-law. In the latter case the spectral break was modelled as the cooling frequency, $\nu_c$, such that the change in slope was fixed at 0.5 \citep{spn98}, and the spectral break was constrained to lie within the UV to X-ray energy range. The physical plausibility of such a change in spectral index is also supported by the analysis of \citet{zwf+11}, who did not constrain the spectral slopes in their SED analysis, but found $\Delta\beta$ to be typically $\sim 0.5$. We considered a broken power-law model to provide a better fit to the afterglow spectral continuum when the F-test returned a null hypothesis probability smaller than 5\%. This relatively high probability threshold was set in order to minimise the chance of any GRBs with an intrinsic broken power-law continuum over our observed bandpass being incorrectly classified as having a single spectral component.

The {\sc xspec} model components {\em phabs} and {\em zphabs} were used, respectively, to fit the Milky Way and host galaxy photoelectric absorption, and {\em zdust} was used to model dust extinction in the Milky Way and host galaxy. The two photoelectric model components translate the soft X-ray absorption resulting from medium weight metals into an equivalent neutral hydrogen column density using a given absorbing cloud metallicity. We assumed this metallicity to be solar, both in the case of the Milky Way and GRB host galaxy, and we used the {\sc xspec} default solar abundances \citep{ag89}. Although these abundances have recently been revised to lower values \citep{ags09}, for typical lines-of-sight within the Milky Way, the \citet{ag89} abundances remain to be good estimates \citep{wat11}. Also, although GRB host galaxy metallicities are typically measured to be sub-solar, the scatter is large and for many host galaxies, metallicity measurements are not available. We therefore chose to assume solar host galaxy metallicities in our analysis both for consistency within our method, as well as to allow us to compare our results with those presented in other published literature, where solar abundances are typically assumed. We fixed the Galactic reddening, $E(B-V)_{Gal}$, and the Galactic neutral hydrogen column density $N_{H,Gal}$, to the values from the Galactic maps of \citet{sfd98} and \citet{kbh+05}, respectively. The Galactic dust extinction was modelled on the mean Milky Way extinction law as parameterised by \citet{pei92}, which has the total-to-selective extinction, $R_V=A_V/E(B-V)$ set to the average value of 3.08 measured in the Milky Way diffuse ISM.

We modelled the dust extinction within the GRB host galaxy on the mean SMC, LMC and Milky Way extinction curves \citep{pei92}, which become increasingly flatter blueward of $\sim 2500$\AA, and have an increasingly pronounced broad extinction feature centred at $\sim$\MWbump\ (see Fig.~\ref{fig:meanextcurve}). These three curves thus provide good coverage of the range in extinction curves observed in the local Universe. To remain consistent with \citet{pei92}, we fixed the total-to-selective extinction of the mean SMC, LMC and Milky Way extinction curves to $R_V=2.93, 3.16$ and 3.08, respectively. This left only the total host galaxy dust reddening along the line-of-sight, $E(B-V)$, free to vary.

\begin{table*}
\caption{Table listing our final sample of 17 GRBs with afterglows best-fit by a power-law SED model and host galaxy \av\ detected at 90\% confidence. \label{tab:subsample}}
\begin{center}
\begin{tabular}{lccccccc}
\hline
GRB & z & Host dust & \av & \nhx & $\beta$ & $\chi ^2$ (dof) & Null hyp. \\
 & & & (mag) & ($10^{21}$~\invsqrcm) & & & prob. \\
\hline\hline
050802 & 1.71 & mw & $0.49\pm0.10$ & $2.36^{+0.75}_{-0.69}$ & $0.72\pm 0.04$ & 73.4 (73) & 0.466 \\
060502A & 1.51 & smc & $0.59^{+0.15}_{-0.12}$ & $3.62^{+1.04}_{-0.93}$ & $0.76^{+0.06}_{-0.05}$ & 23.9 (29) & 0.736 \\
060607A & 3.082 & smc & $0.08\pm 0.04$ & $12.86^{+4.09}_{-3.53}$ & $0.52\pm 0.02$ & 29.3 (31) & 0.555 \\
060904B & 0.703 & lmc & $0.15\pm 0.04$ & $2.98^{+0.62}_{-0.54}$ & $0.85\pm 0.01$ & 36.5 (29) & 0.158 \\
060912 & 0.937 & mw & $0.46^{+0.23}_{-0.22}$ & $2.96^{+0.70}_{-0.63}$ & $0.85^{+0.07}_{-0.06}$ & 16.0 (21) & 0.770 \\
061007 & 1.262 & smc & $0.44\pm 0.01$ & $6.00\pm 0.21$ & $0.89\pm 0.01$ & 285.5 (289) & 0.548 \\
070318 & 0.836 & smc & $0.50\pm 0.04$ & $8.07^{+0.60}_{-0.54}$ & $1.14\pm 0.02$ & 63.8 (59) & 0.312 \\
070810A & 2.17 & mw & $0.61^{+0.24}_{-0.21}$ & $7.89^{+2.26}_{-2.02}$ & $1.01^{+0.09}_{-0.08}$ & 18.3 (23) & 0.742 \\
071112C & 0.823 & smc & $0.20^{+0.05}_{-0.04}$ & $0.82^{+0.89}_{-0.72}$ & $0.58\pm 0.02$ & 23.7 (17) & 0.128 \\
080319B & 0.937 & smc & $0.06^{+0.03}_{-0.02}$ & $1.13^{+0.42}_{-0.37}$ & $0.72\pm 0.01$ & 15.1 (14) & 0.372 \\
080319C & 1.95 & smc & $0.71^{+0.08}_{-0.07}$ & $8.73^{+1.12}_{-1.03}$ & $0.58\pm 0.04$ & 99.7 (98) & 0.432 \\
080710 & 0.845 & smc & $0.03\pm 0.01$ & $0.49^{+0.29}_{-0.26}$ & $0.87\pm 0.01$ & 39.2 (37) & 0.370 \\
080928 & 1.692 & smc & $0.24\pm 0.06$ & $3.54^{+1.14}_{-1.01}$ & $1.09^{+0.04}_{-0.05}$ & 33.8 (30) & 0.288 \\
081008 & 1.968 & lmc & $0.29\pm 0.07$ & $7.60^{+1.82}_{-1.62}$ & $1.13^{+0.04}_{-0.03}$ & 13.2 (23) & 0.947 \\
090927 & 1.37 & mw & $0.40^{+0.18}_{-0.15}$ & $2.20^{+1.52}_{-1.25}$ & $0.88\pm 0.08$ & 2.2 (9) & 0.989 \\
091020 & 1.71 & lmc & $0.86^{+0.10}_{-0.09}$ & $5.12^{+0.96}_{-0.88}$ & $1.09^{+0.05}_{-0.03}$ & 40.0 (35) & 0.256 \\
091208B & 1.063 & lmc & $0.95^{+0.22}_{-0.20}$ & $7.81^{+1.40}_{-1.21}$ & $0.90^{+0.09}_{-0.07}$ & 20.9 (30) & 0.893 \\
\hline
\end{tabular}
\end{center}
Columns (2) and (3) give the GRB redshift and the best-fit host galaxy extinction law; either smc, lmc or mw to refer to the SMC, LMC or Milky Way extinction model fit, and columns (4) to (6) are the best-fit host galaxy \av, host galaxy \nh, and GRB energy spectral index, $\beta$. The last two columns give the $\chi ^2$ and degrees of freedom (dof), and the null hypothesis probability of the best-fit solution.
\end{table*}

Lyman-series absorption in the 912--1215\AA\ rest-frame wavelength range from neutral hydrogen in the intergalactic medium (IGM) was modelled using results from \citep{mad95}, and the uncertainty associated to this was added in quadrature to the photometric error of any optical data at rest-frame wavelengths blueward of Ly$\alpha$. The uncertainty on the IGM absorption stems from statistical fluctuations in the number of absorbing clouds along the line-of-sight, and it is thus a function of redshift, which in our sample reached up to 20\%.

\section{The GRB mean extinction curve}
\label{sec:method}
To study the dust extinction curve within GRB host galaxies, we wish first to reduce our sample of 49 GRBs down to those that have both a significant host galaxy dust extinction signature imprinted on their SED and that have a single power-law NIR to X-ray spectral component. This latter requirement is applied because of the degeneracy between the best-fit optical spectral index and the steepness of the extinction curve fitted to the SED. Over-predicting the steepness of the UV extinction curve when modelling the afterglow SED would result in a best-fit optical spectral index that is flatter than its true value. Such an over-prediction would either introduce an artificial break in the afterglow broadband continuum, or move the location of a true break to smaller energies. In a similar regard, an under-estimation of the extinction curve steepness would result in a steeper best-fit spectral index than the true value. However, whereas in the former case an erroneous best-fit spectral index may still provide an acceptable fit to the SED by shifting the position of the cooling break, a steeper spectral index would be inconsistent with the X-ray spectral index in the context of our SED synchrotron emission model. An under-estimation of the extinction curve steepness would thus result in a poor fit to the data in our SED analysis. By only selecting those GRBs with a single intrinsic NIR to X-ray afterglow spectral component, we remove this degeneracy, and are thus left with a sample where we know our SED analysis to be accurate. This final level of filtering, by definition, removes those GRBs from our sample that have host galaxy extinction curves with very different slopes to those that we have used in our analysis, and we shall return to this point in section~\ref{sssec:fltbias}.

Of the 49 GRBs analysed, 25 did not require a spectral break between the X-ray and optical wavelength ranges. As a consistency check on our SED analysis, we also verified that the temporal decay slope in the X-ray and optical bands was the same within $1\sigma$ errors at the epoch of the instantaneous SED. 36 of the 49 GRBs had a host galaxy \av\ measured to be greater than zero at the 90\% confidence, and 17 of these were best-fit by a single spectral component, and thus made our final sample. The NIR to X-ray SEDs of these 17 GRBs are plotted in Fig.~\ref{fig:fnlSEDs} with the best-fit FM90 model overplotted, which we described in the next section. The SEDs together with the best-fit models of the 32 GRBs that did not enter our final sample are plotted in Fig.~\ref{fig:SEDs}. In Table~\ref{tab:subsample} we summarise the best-fit model results for each of the 17 GRBs in our prime sample.

\begin{table*}
\caption{Best-fit results from our simultaneous SED analysis.\label{tab:fitres}}
\begin{center}
\begin{tabular}{lccccccccc}
\hline
Model & $c_1$ & $c_2$ & $c_3$ & $c_4$ & $\gamma$ & $x_c$ & $\chi2/dof$ & Null hyp.\\
& $=2.09-2.84c_2$ & & & & & & & prob.\\ 
\hline\hline
{\bf FM90} & & & & & & \\
{\em All data} & -2.66 & $1.67^{+0.81}_{-1.51}$ & $0.00^{+0.67}_{-0.00}$ & $0.43^{+0.14}_{-0.11}$ & 0.92 & 4.59 & 819.9/846 & 0.734 \\
{\em $\lambda_{UV}>121$nm} & -1.64 & $1.31^{+0.79}_{-1.20}$ & $0.14^{+0.57}_{-0.18}$ & $1.80^{+1.29}_{-0.64}$ & 0.92 & 4.69 & 813.8/814 & 0.495 \\
\hline
{\bf SMC} & & & & & & & \\
{\em All data} & {\em-4.47$\pm$0.19} & {\em 2.35$\pm$0.18} & {\em 0.08$\pm$0.01} & {\em -0.22$\pm$0.01} & {\em 1.00} & {\em 4.60} & 815.0/812 & 0.464 \\
{\em $\lambda_{UV}>121$nm} & & & & & & & 786.5/781 & 0.438 \\
\hline
{\bf LMC} & & & & & & & \\
{\em All data} & {\em -2.16$\pm$0.36} & {\em 1.31$\pm$0.08} & {\em 1.92$\pm$0.23} & {\em 0.42$\pm$0.08} & {\em 1.05$\pm$0.07} & {\em 4.63$\pm$0.01} & 818.7/812 & 0.428 \\
{\em $\lambda_{UV}>121$nm} & & & & & & & 795.4/781 & 0.352 \\
\hline
{\bf Milky Way} & & & & & & & \\
{\em All data} & {\em 0.12$\pm$0.11} & {\em 0.63$\pm$0.04} & {\em 3.26$\pm$0.11} & {\em 0.41$\pm$0.02} & {\em 0.96$\pm$0.01} & {\rm 4.60$\pm$0.002} & 1333.1/812 & $5.5e^{-28}$\\
{\em $\lambda_{UV}>121$nm} & & & & & & & 1276.3/781 & $1.7e^{-26}$\\
\hline
\end{tabular}
\end{center}
NOTE- Table lists results from both the analysis that used all available afterglow SED data, and from the analysis where data covering the Lyman-forest were excluded (i.e. no UV data at wavelengths $\lambda <121.5$~nm) . The best-fit \citet{fm90} (FM90) parameterisations to the SMC, LMC and Milky Way models from \citet{cws+03} are shown in italic as a comparison to the best-fit FM90 fits to our data.
\end{table*}

To investigate the mean shape of the extinction curve in the host galaxies of our sample of 17 GRBs, we applied a simultaneous fit to all 17 NIR to X-ray broadband SEDs using similar models as those described in section~\ref{ssec:SEDs}. However, in addition to the SMC, LMC and Milky Way extinction laws previously fitted to the afterglow SEDs, we also fitted the more general extinction curve model developed by \citet{fm90} to our data, which we shall refer to as the FM90 model. This latter extinction curve model has eight free parameters that are able to fit variations in the NIR and UV curvature and in the prominence of the \MWbump\ feature observed along lines-of-sight within local and extragalactic galaxies. The large number of degrees of freedom in the FM90 model did not make it a feasible model to use in the individual afterglow SED analysis. However, the large number of data points in our simultaneous fit provide sufficient constraint for the FM90 model to be fitted to the data. Additionally, the widespread use of this model \citep[e.g.][]{cws+03,gcm+03,efh+09,pmu+11} allows us to compare our best-fit results to the dust extinction properties of other local and extragalactic environments.

The eight free parameters of the FM90 model are made up of the total galaxy dust reddening, $E(B-V)$, the total-to-selective dust extinction, \rv, and six further coefficients. These six coefficients define the linear extinction component underlying the UV range ($c_1$ and $c_2$), the height ($c_3$), width ($\gamma$), and central wavelength ($x_c$) of the \MWbump\ bump, which is modelled with a Lorentzian-like {\em Drude} profile ($D$($x$;$\gamma$,$x_c$)), and the far-UV curvature ($c_4$). The full equation is given in \citet{fm90}, and here we just provide the unexpanded formulism, for simplicity:
\begin{eqnarray}
\label{eq:fm90}
\lefteqn{A_\lambda=E(B-V)\times}\\
&&~~~[R_V+c_1+c_2\lambda^{-1}+c_3D(\lambda^{-1})+c_4F(\lambda^{-1})]\nonumber
\end{eqnarray}
The UV intercept, $c_1$, and the UV slope, $c_2$, are tightly correlated, with the most recent analysis showing a linear dependence of $c_1 = 2.09 - 2.89c_2$ \citep{fm07}. Furthermore, the central wavelength of the \MWbump\ bump does not vary significantly. We therefore fix $x_c$ to the mean observed value of $4.592~\mu m^{-1}$, and tie $c_1$ to $c_2$, as described above, leaving us with six free parameters. Given the typically weak signature of a bump in the extinction law of GRB host galaxies together with the low spectral resolution available from our broadband data, the constraint on the remaining two bump parameters (i.e. $c_3$ and $\gamma$) is small, and we therefore also fixed the width of the bump, $\gamma$, to the median Milky Way value of $\gamma =0.922$ \citep{fm07}. This allows for a more realistic measurement of the bump strength or an upper limit, provided by the parameter $c_3$.

In its original form the FM90 model is only valid up to wavelengths $\lambda <2700$\AA\ (rest-frame). Furthermore, most of the GRBs in our sample do not have sufficient data at rest-frame NIR wavelengths to derive the total-to-selective extinction, which is given by $R_V=1.10E(V-K)/E(B-V)$ \citep{mn82}. We therefore appended the mean LMC extinction curve to the FM90 analytic model to cover the extinction at wavelengths $\lambda >2700$\AA, and fixed \rv\ to the mean LMC value, \rv=3.16 \citep{pei92}. We chose the LMC curve because it has a continuum that is mid-way between that observed in the SMC and Milky Way. Nevertheless, all well-observed extinction curves in the local Universe are comparable redward of 2500\AA, and the specific choice of these appended to the \citet{fm90} model at $\lambda > 2700$\AA\ has only a marginal effect on our results.

In our simultaneous SED analysis, all dust extinction models fitted had the strength of the host galaxy reddening, E(B-V), as a free parameter for each GRB, but all other dust parameters were tied between GRBs. The spectral index of the continuum, $\beta$, and the amount of host metal absorption, \nh, were also left free to vary for each GRB.

\section{Results}
\label{sec:results}
The simultaneous best-fit extinction curves fitted to our data are plotted in Fig.~\ref{fig:meanextcurve}, all normalised at $\lambda = 3000$\AA. The shaded grey area is the $90\%$ confidence region to the best-fit FM90 model. As a visual reference we also plot in Fig.~\ref{fig:smplfits} all 17 SEDs from our sample together with the best-fit model and residuals from each of the four host galaxy extinction curve models fitted. A summary of our best-fit results is given in Table~\ref{tab:fitres}, and to place our best-fit FM90 coefficients in context, in this table we also list the FM90 best-fit coefficients to the average SMC, LMC and Milky Way extinction laws \citep[][and references therein]{cws+03}.

We find that the Milky Way mean extinction law cannot adequately fit the host galaxy extinction properties of our sample of GRB afterglows, and is rejected at 99.99\% confidence (see Table~\ref{tab:fitres}). This is also illustrated by the inconsistency between the best-fit FM90 extinction curve and the Milky Way mean extinction curve plotted in Fig.~\ref{fig:meanextcurve}, where the latter is significantly flatter and has a more pronounced \MWbump\ bump. In the bottom right-hand panel of Fig.~\ref{fig:smplfits}, the residuals from the Milky Way model fitted to the data also show a clear excess in the data-to-model ratio at the location of the Milky-Way \MWbump\ bump. Both the SMC and LMC mean extinction laws, on the other hand, provide good fits to the data. The best-fit FM90 UV continuum is in closer agreement to the LMC rather than the SMC UV extinction slope, although at the location of the \MWbump\ bump, the best-fit FM90 model is closer to the featureless SMC curve. Nevetherless, the $90\%$ confidence region of the FM90 model plotted in Fig.~\ref{fig:meanextcurve} shows that the presence of a weak \MWbump\ bump cannot be ruled out by our data.

We used the best-fit FM90 extinction curve from our simultaneous SED analysis to re-fit the SEDs of all 49 GRBs that were in our initial sample. In these fits, all FM90 parameters were fixed to their best-fit value from our simultaneous analysis, and only the host galaxy reddening, $E(B-V)$, was left free to vary. This host extinction curve model provided an acceptable fit to all 49 GRB afterglow SEDs, and compared to the SMC, LMC and Milky Way models, improved the fit for 65-70\% of the GRBs with host galaxy dust extinction measured at 90\% confidence. For the remaining 30-35\%, the goodness of fit worsened by a median of  $\langle \Delta\chi^2\rangle=1$ relative to the SMC fit, but by $\langle\Delta\chi^2\rangle=5$ for the Milky Way fit. By allowing the height of the 2175\AA\ bump to vary, the fraction of GRBs better fit by the FM90 model rose to $\sim 80$\%. This indicates that for {\rm at least} 10--15\% of the extinguished GRBs in our sample, the host galaxy extinction curve has a 2175\AA\ feature. This is in good agreement with the results from \citet{zwf+11}, who studied the extinction curve for a sample of 41 GRB afterglows using combined, broadband photometric and spectroscopic data. In their analysis, they found $\sim 85$\% of their extinguished sample to be best-fit by SMC-type extinction, and $\sim 10$\% to be best-fit by an extinction curve model with a 2175\AA\ feature.

\section{Discussion}
\label{sec:disc}
In our analysis on the host galaxy dust extinction imprinted on the UV through to NIR afterglows of a prime sample of 17 GRBs, we find the best-fit model to be consistent with both the mean SMC and LMC extinction curve. This is in agreement with the results from several previous studies on the dust extinction properties of GRB hosts \citep[e.g.][]{sfa+04,kkz06,sww+07,smp+07,hlp+08,spo+10}, although contradicts the findings of \citet{clw06} and \citet{llw08}, who found GRB hosts to have a considerably flatter extinction curve, comparable to or flatter than the Milky Way mean extinction curve. In \citet{clw06} and \citet{llw08} the optical intrinsic spectrum, $\beta_{\rm opt}$, was inferred indirectly by using the closure relations defined by synchrotron emission theory \citep{zm04}. In \citet{clw06}, $\beta_{\rm opt}$ was calculated from the optical temporal decay index, $\alpha_{\rm opt}$. \citet{llw08}, on the other hand, used a sample of GRBs with consistent X-ray and optical afterglow temporal indices, $\alpha_{\rm opt}$ and $\alpha_{\rm x}$, and assumed the optical spectral index, $\beta_{\rm opt}$, to be the same as the measured X-ray spectral index, $\beta_{\rm x}$. However, caution must be taken in such analysis, given the large fraction of GRBs that have optical and X-ray afterglow data that are inconsistent with the simplest fireball scenarios \citep[e.g][]{pmb+06,ebp+09,ngg+10}, and although $\beta_{\rm opt}=\beta_{\rm x}$ implies $\alpha_{\rm opt}=\alpha_{\rm x}$, the opposite does not necessarily apply \citep[e.g.][]{fkg+11}. Further complications arise from the degeneracy that there exists between the X-ray spectral index and the amount of soft X-ray absorption. For example, an error of 0.1 on the X-ray photon spectral index, $\Gamma_{\rm x}$, would introduce an uncertainty of $\sim 0.5$ magnitude in the $V$-band flux, and a break between the X-ray and optical emission corresponding to the cooling frequency ($\Gamma_{\rm opt}=\Gamma_{\rm x}-0.5$) would reduce the predicted flux in the optical by a factor of $\sim 10$.

The shape of our best-fit GRB host extinction curve strongly contrast with the extinction curves of the highly extinguished GRBs 070802 and 080607 measured in \citet{efh+09} and \citet{pmu+11} respectively, and GRB~080605 and GRB~080805 as measured by \citet{zwf+11}, all of which are flatter and have a clear 2175\AA\ feature. None of these GRBs were included in our sample due to the faintness of their observed optical afterglows. To investigate the possible origins for these differences in host extinction curves we consider first the systematic uncertainties present in our analysis, as well as the selection effects in our sample. We then compare properties of those GRBs in our sample with those of GRBs with distinct host dust extinction properties.

\subsection{Systematic biases}
There are two principle areas where systematics may be introduced, and this is i) in our modelling of the Galactic dust extinction and soft X-ray absorption, and ii) in the effect of host galaxy neutral hydrogen on our UV and optical data.

\subsubsection{Milky Way dust extinction}
The uncertainty in the $N_{H,Gal}$ and $A_{V,Gal}$ can be up to a factor of two when the line-of-sight is close to the Galactic plane. However, probably as a consequence of our selection criteria requiring a GRB peak $v$-band afterglow $v < 19$, only three of the initial 49 GRBs analysed have a Galactic visual extinction $A_V > 0.5$, at which point the uncertainty in the Galactic reddening becomes significant. Only one of these three GRBs entered our final sample. Furthermore, due to the general inverse wavelength dependence of extinction, any extinction in the observed optical bands is dominated by the uncertainty in the much larger UV rest-frame extinction. Therefore, although the SED analysis of these three more heavily Milky-Way extinguished GRB afterglows may be uncertain, this should only introduce scatter rather than alter our results. In the case of the Galactic neutral hydrogen column, $N_{H,Gal}$, the uncertainty is typically less than 10\%, which is small relative to the uncertainty on the best-fit host galaxy soft X-ray absorption. 

Another source of uncertainty is in the Milky Way extinction law, which varies along different lines-of-sight as a function of \rv\ \citep{ccm89}. The scatter around the mean value of \rv=3.1 is, however, small, with a root mean square of 0.1 \citep{sw75}. Furthermore, larger values of \rv\ are seen along lines-of-sight that cross dense regions of dust, and are therefore not very common. This is especially true for our sample, which is biased against heavily extinguished lines-of-sight. We therefore do not expect variations in the Galactic extinction curve along the lines-of-sight to our sample of GRBs to have a significant affect on our final results.

\begin{figure}
\includegraphics[width=0.37\textwidth,angle=-90]{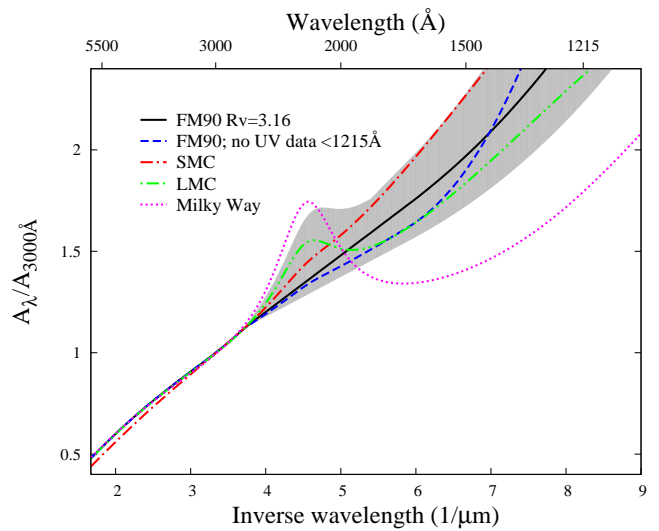}
\caption{Best-fit mean GRB host extinction law as derived from our simultaneous SED fits normalised at $A_{\rm 3000\AA}$ and corresponding 90\% confidence region (grey). The best-fit FM90 model is plotted in black (solid), and the best-fit model to the SEDs with UV data blueward of 1215\AA\ omitted is shown in blue (long-dashed). Also shown for comparison are the mean SMC (red, dot-dash), LMC (green, dot-dot-dash) and Milky Way (pink, dotted) extinction laws, using the parameterisation from \citep{pei92}.}
\label{fig:meanextcurve}
\end{figure}

\subsubsection{Host neutral hydrogen absorption}
A factor that needs to be addressed before discussing our results is the effect of gas absorption at UV wavelengths within the host galaxy, which is unaccounted for in our SED modelling. The presence of a strong damped Lyman-$\alpha$ (DLA) system within the GRB host galaxy, which can contain column densities of neutral gas as large as $N_H\sim 10^{23}$~\invsqrcm\ \citep[e.g. GRB~050401 and GRB~060926; ][]{fjp+09}, may systematically bias our dust extinction measurements at rest-frame wavelengths 1150--1250\AA. If indeed significant, then this neutral hydrogen absorption within the host galaxy may be incorrectly attributed to dust

\onecolumn
\begin{figure}
\begin{center}
\includegraphics[height=0.93\textwidth,angle=-90]{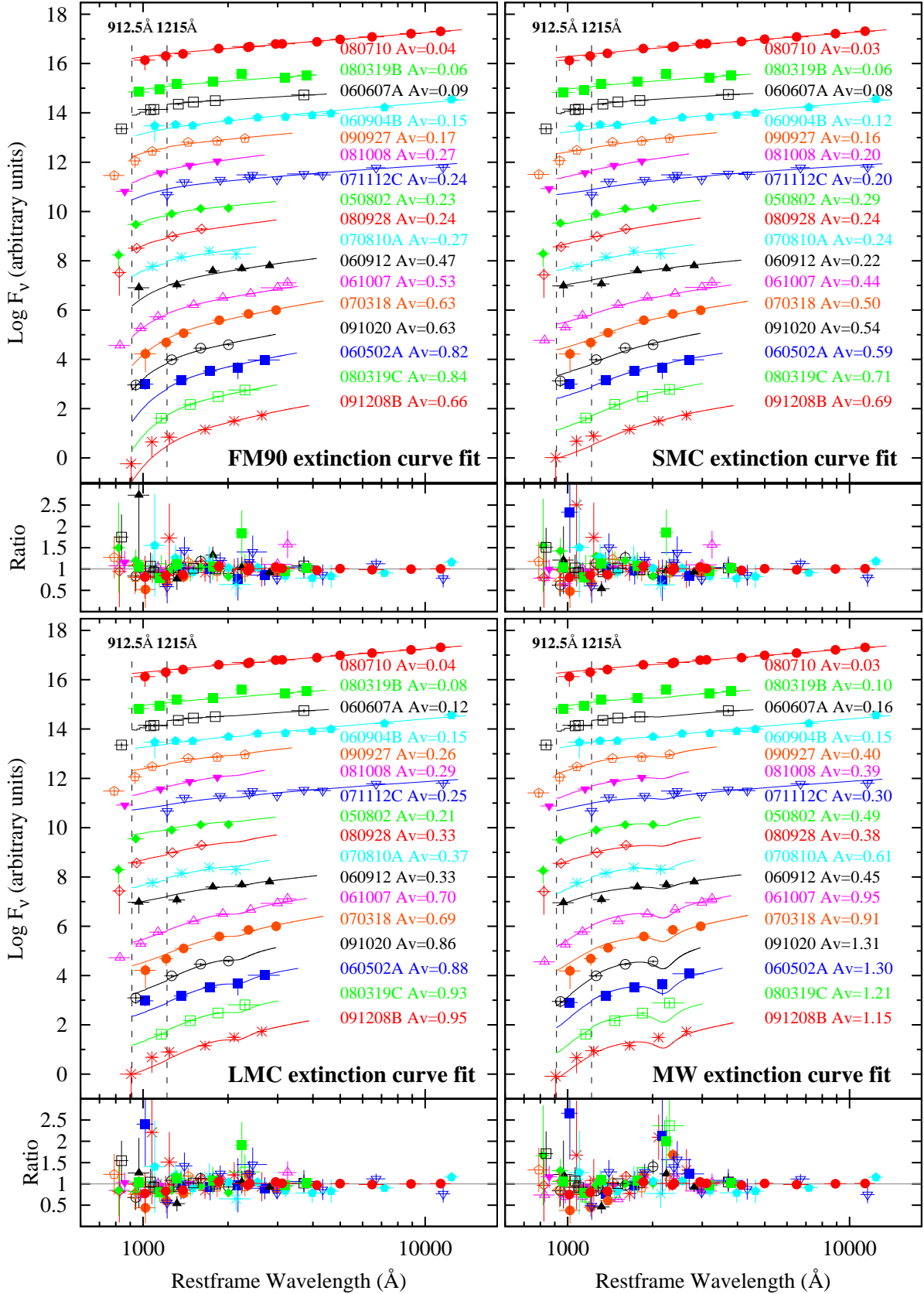}
\end{center}
\caption{UV/optical/NIR afterglow SEDs of the 17 GRBs in our sample plotted in the rest-frame. Overplotted are the best-fit models from the simultaneous fits, where each of the four figures correspond to a different host galaxy extinction curve model fitted to the data. The four extinction curves fitted are our best-fit FM90 extinction curve (top left; best-fit coefficients given in Table~\ref{tab:fitres}), the mean SMC (top right), mean LMC (bottom left), and mean Milky Way (bottom right) extinction curves, where the latter three extinction curves use the parameterisation from \citet{pei92}. In all panels the y-axis is arbitrary and the afterglow SEDs are mostly ordered by afterglow host galaxy extinction with the most extinguished at the bottom, and the least plotted at the top. In all four panels the dashed lines indicate the Lyman-break at 912.5\AA, and the location at 1215\AA, blueward of which the Lyman-forest presents a further absorption component.}\label{fig:smplfits}
\end{figure}

\twocolumn
\noindent
extinction, and thus result in an apparent steepening of the best-fit extinction curve in the rest-frame UV. There is also uncertainty associated with neutral hydrogen absorption within the IGM. However, we model this on the {\em average} line-of-sight out to a given redshift, and our IGM absorption model will, therefore, not introduced any systematic effects. To investigate the effect that neutral host gas has on our results we re-ran our simultaneous afterglow SED fits, but this time using only UV data redward of the Lyman forest ($\lambda > 1215$\AA\ rest-frame). The best-fit results from these fits are summarised in Table~\ref{tab:fitres}, and the best-fit extinction curve is also plotted in Fig.~\ref{fig:meanextcurve} (blue; long-dashed). The best-fit coefficients that describe the underlying UV continuum ($c_1$ and $c_2$), and the height of the \MWbump\ bump ($c_3$), are not greatly affected by the removal of UV data blueward of $1215$\AA, and remain consistent within their $1\sigma$ errors. The new best-fit value of $c_4$, which describes the far-UV extinction curvature at wavelengths $\lambda<1700$\AA, shows greater variation, increasing from $0.43^{+0.13}_{-0.11}$ in our full data-set analysis, to $1.79^{+1.29}_{-0.64}$ when data points that lie within the Lyman forest were removed. Nevertheless, the values remain consistent within $2\sigma$.

Given that the removal of those data points blueward of $\lambda = 1215$\AA\ reduces the constraint on the far-UV curvature by over 50~\%\footnotemark[5], it is not surprising that the $c_4$ parameter shows the greatest variation in our fits with or without UV data at $\lambda < 1215$\AA.
\footnotetext[5]{Of the 46 data points at rest-frame $\lambda<1700$\AA, where the $c_4$ coefficient becomes relevant, 25 lie within the Lyman forest (i.e. rest-frame $\lambda < 1215$\AA)} 
Furthermore, the fact that $c_4$ increases in the latter fits suggests that neutral gas absorption within the host galaxy does not have a significant effect on our results. If the UV data blueward of 1215\AA\ had been significantly affected by host galaxy neutral hydrogen absorption, we would expect the omission of these data in our SED analysis to have yielded a smaller best-fit value of $c_4$ relative to our initial analysis (i.e. the far-UV slope of the extinction law to have flattened), which is not the case. Furthermore, the best-fit extinction curve to our reduced data set lies within the $3\sigma$ confidence region of our best-fit results to the full data set (Fig.~\ref{fig:meanextcurve}). This therefore suggests that our results are not significantly affected by additional absorption of the UV afterglow from host galaxy neutral hydrogen.

\begin{figure}
\begin{center}
\includegraphics[height=0.40\textwidth]{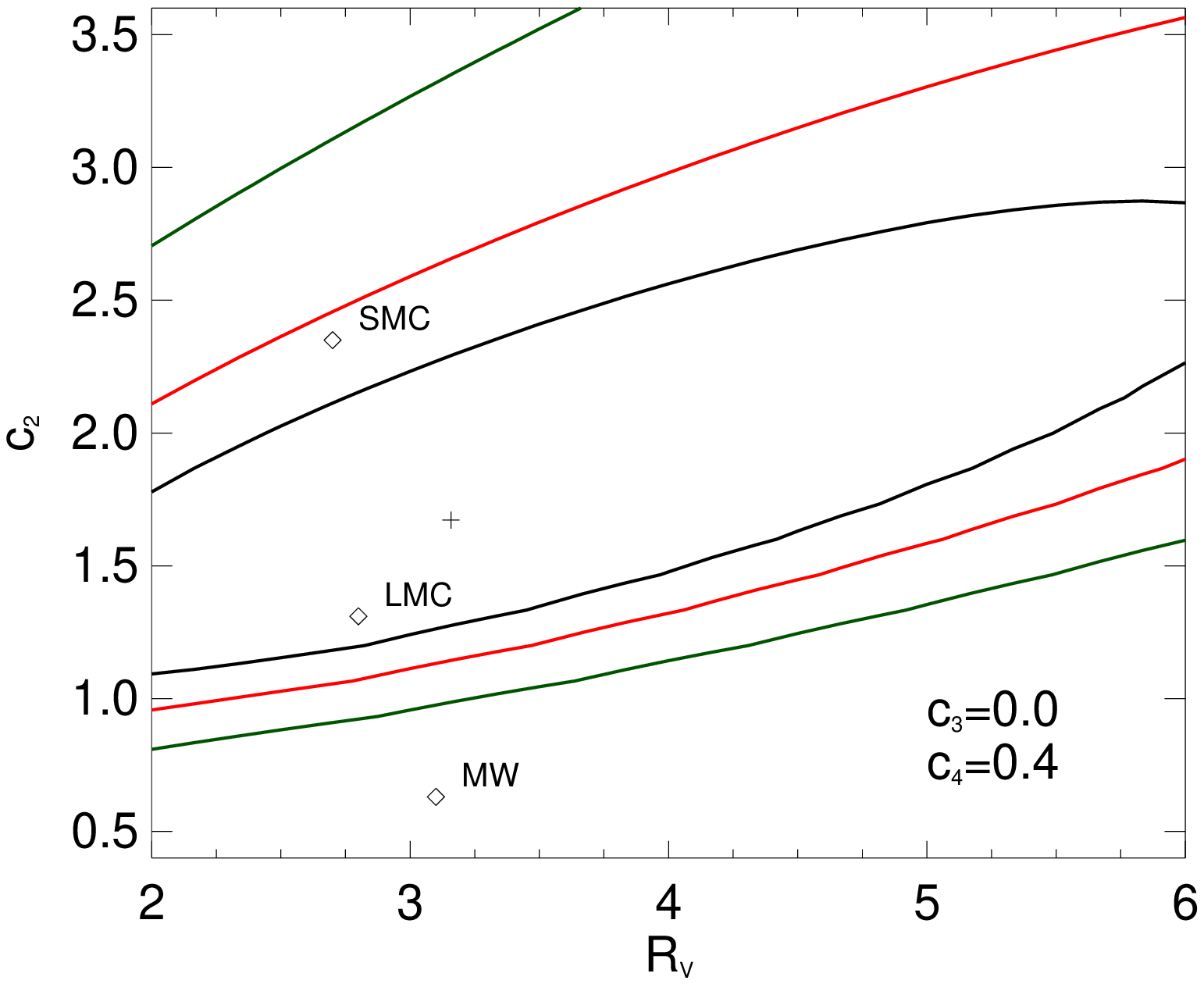}
\includegraphics[height=0.40\textwidth]{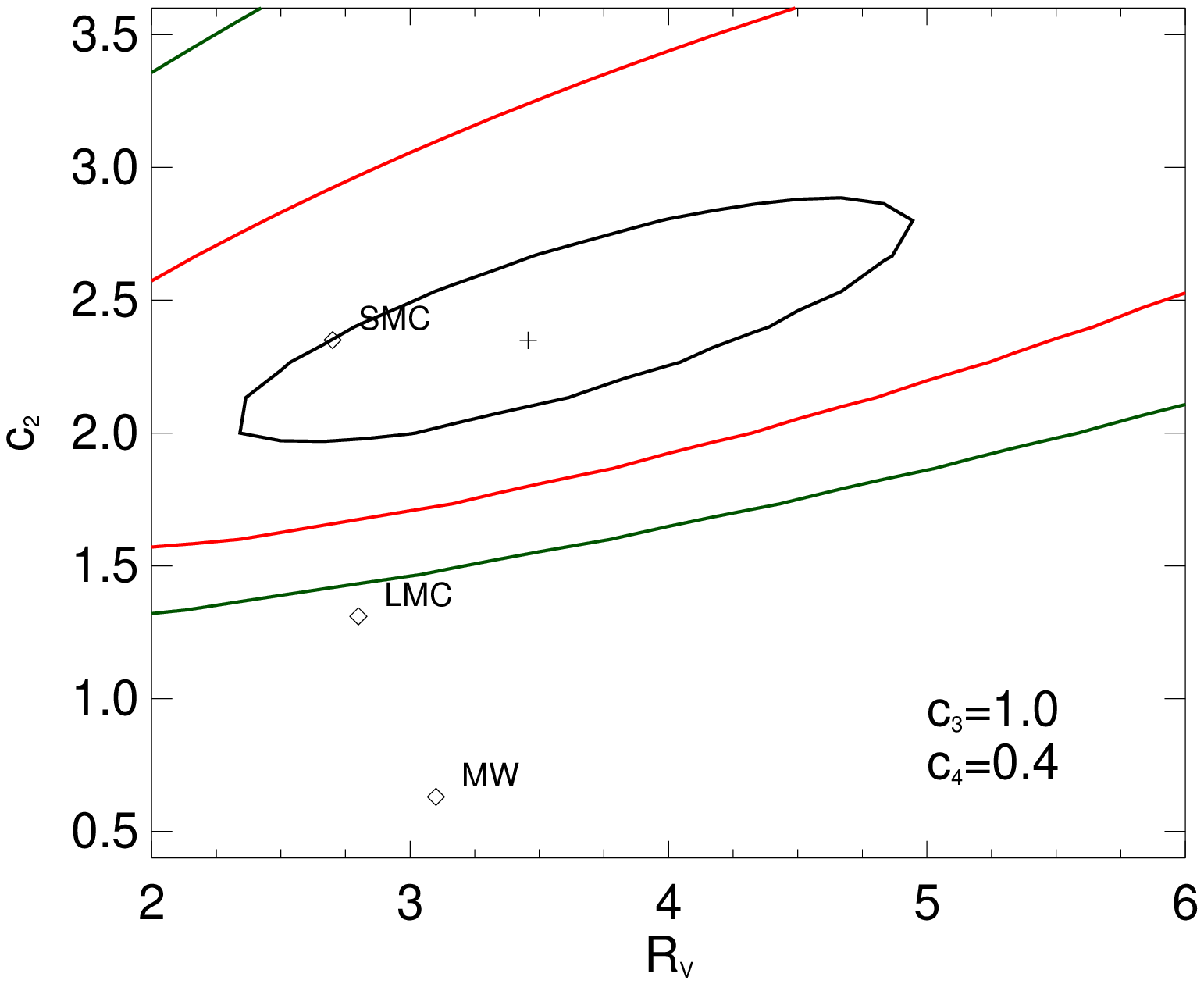}
\end{center}
\caption{2-dimensional contour plot of $c_2$ against \rv\ for $c_3=0.0$ (top) and $c_3=1.0$ (bottom), and $c_4=0.4$ in both cases. In both panels, the inner (black), middle (red) and outer (green) contours correspond to 68.3\%, 90\% and 99\% confidence intervals, respectively. The best-fit $c_2$ and \rv\ values are indicated by a black cross, and the corresponding [\rv,$c_2$] pairs for the mean SMC, LMC and Milky Way (MW) extinction curve are also shown.}\label{fig:rvc2cont_c3}
\end{figure}

\begin{tiny}
\begin{table*}
\caption{Best-fit results from FM90 model fits to the afterglow SEDs of 24 GRBs, all of which were all best-fit by a broken-power model in our initial SED analysis (see Table~\ref{tab:GRBsamp}).\label{tab:refits}}
\begin{center}
\begin{tabular}{l|cll|lll|c}
\hline
GRB & Local-dust$^a$ & & & \multicolumn{3}{c}{Power Law} & $\alpha_x\sim\alpha_o$\\
 & model & $E(B-V)$ & $\chi^2$/dof & $c_1$ & $E(B-V)$ & $\chi^2$/dof & (Y/N) \\
\hline\hline
050319  & mw & $0.10^{+0.03}_{-0.05}$ & 58/78 & $3.60^{+0.37}_{-0.32}$ & $0.44^{+0.11}_{-0.10}$ & 62/78 & Y \\
050820A & lmc & $0.06\pm 0.01$ & 177/143 & $2.95\pm 0.25$ & $0.23\pm 0.05$ & 190/143 & N \\
060206  & mw & $<0.08$ & 61/59 & $4.39^{+0.61}_{-10.9}$ & $<0.03$ & 84/59 & N \\
060418  & smc & $0.045^{+0.004}_{-0.007}$ & 19/23 & $6.49^{+5.84}_{-0.01}$ & $<0.01$ & 29/23 & N \\
060729  & smc & $0.05\pm 0.01$ & 176/191 & $2.78^{+2.22}_{-9.28}$ & $<0.15$ & 180/191 & Y \\
{\bf 061121}  & lmc & $0.17^{+0.02}_{-0.03}$ & 87/93 & $2.65^{+0.10}_{-0.46}$ & $0.73^{+0.11}_{-0.12}$ & 86/93 & N \\
{\bf 061126}  & lmc & $0.18^{+0.02}_{-0.03}$ & 170/140 & $1.92^{+0.36}_{-0.46}$ & $0.36\pm 0.05$ & 168/140 & N \\
070110  & smc & $<0.07$ & 129/112 & $3.39^{+0.85}_{-0.05}$ & $0.82^{+0.17}_{-0.15}$ & 137/112 & Y \\
080210  & lmc & $0.08\pm 0.01$ & 23/25& $-1.22^{+1.17}_{-1.83}$ & $0.09\pm 0.02$ & 35/25 & N \\
080411  & mw & $<0.02$ & 81/83 & $^b$ & $0.00$ & 216/83 & Y \\
{\bf 080413B} & mw & $0.27^{+0.04}_{-0.06}$ & 78/69 & $2.94^{+0.34}_{-0.39}$ & $0.60^{+0.07}_{-0.10}$ & 75/69 & Y \\
080430  & smc & $<0.07$ & 38/40 & $3.91^{+0.86}_{-0.36}$ & $0.21^{+0.09}_{-0.08}$ & 52/40 & Y \\
{\bf 080721} & mw & $0.13\pm 0.05$ & 168/153 & $3.91^{+0.31}_{-0.22}$ & $0.77\pm 0.06$ & 168/153 & N \\
080804  & lmc & $0.033\pm 0.004$ & 38/42 & $^b$ & $<0.01$ & 92/42 & N \\
080916A & smc & $0.21\pm 0.05$ & 18/26 & $1.24^{+0.66}_{-1.07}$ & $0.64^{+0.18}_{-0.17}$ & 19/26 & N \\
081121   & smc & $0.023^{+0.005}_{-0.004}$ & 52/47 & $^b$ & $<0.01$ & 92/47 & N \\
{\bf 081203A} & mw & $0.09\pm 0.01$ & 76/80 & $-0.76^{+0.14}_{-0.43}$ & $0.23\pm 0.05$ & 61/80 & Y \\
081222  & mw & $0.012^{+0.006}_{-0.004}$ & 34/46 & $^b$ & $<0.01$ & 155/46 & N \\
090102  & mw & $0.13\pm 0.01$ & 22/27 & $0.22^{+0.66}_{-0.83}$ & $0.16\pm 0.02$ & 35/27 & N \\
090424  & mw & $0.37\pm 0.03$ & 92/83 & $1.94^{+0.45}_{-0.62}$ & $0.52\pm 0.08$ & 95/83 & Y \\
090618  & smc & $0.053^{+0.003}_{-0.004}$ & 24/18 & $<3.7$ & $<0.02$ & 36/18 & Y \\
091018  & smc & $0.033\pm 0.005$ & 55/52 & $^b$ & $0.00$ & 101/52 & N \\
091029  & lmc & $0.02\pm 0.01$ & 23/23 & $^b$ & $<0.02$ & 58/23 & Y \\
091127  & lmc & $<0.01$ & 186/173 & $^b$ & $0.00$ & 326/173 & N \\
\hline
\end{tabular}
\end{center}

NOTE- All FM90 model parameters were fixed to the best-fit values from the simultaneous SED analysis (see Table~\ref{tab:fitres}) apart from the coefficient $c_1$, which determines the slope of the extinction curve. Column 8 indicates whether the optical and X-ray light curve decay rates are consistent (Y) or not (N) at epoch of the corresponding GRB afterglow SED. Those GRBs better fit by a single power-law spectral component and an FM90 host extinction curve are highlighted in bold (see text for details).\\
$^a$ The best-fit local dust model when using the broken power-law model, chosen from the mean SMC (smc), LMC (lmc) and Milky Way (mw) extinction laws.\\
$^b$ $c_1$ is unconstrained\\
\end{table*}
\end{tiny}

\subsection{Selection effects}
\subsubsection{How constrained is the extinction curve slope?}
The optical slope of extinction curves is commonly parameterised by the total-to-selective extinction parameter \rv$=A_{\rm V}/E(B-V)$, whereby for a fixed \av, a larger \rv\ would imply a smaller amount of reddening, $E(B-V)$, and thus a flatter extinction curve. As mentioned in section~\ref{sec:method}, our dataset does not have sufficient optical and NIR rest-frame coverage to constrain \rv. Nevertheless, our UV data {\em do} provide constraints on the value of \rv\ {\em relative} to the two coefficients, $c_1$ and $c_2$, which describe the continuum of the extinction curve at UV wavelengths. When we substitute $c_1$ for $2.09 - 2.84c_2$, the FM90 parameterisation that we have adopted (see Eq.~\ref{eq:fm90}) then reduces to
\begin{eqnarray}
\label{eq:modfm90}
 \lefteqn{A_\lambda=E(B-V)\times}\\
&&~~~[R_V+2.09-2.84c_2+c_2/\lambda+c_3D(\lambda^{-1})+c_4F(\lambda^{-1})]\nonumber
\end{eqnarray}
for $\lambda$ in units of microns, and for $\lambda<0.27\mu$m, and where the other terms are as defined in section~\ref{sec:method}. It is clear from Eq.~\ref{eq:modfm90} that the parameters $R_V$ and $c_2$ both contribute to the wavelength independent term. However \rv\ and $c_2$ are not completely degenerate because the $c_2$ parameter also appears in the inverse wavelength dependent term. We have used {\sc xspec} to determine what parts of $R_V-c_2$ parameter space can provide a statistically acceptable fit to our GRB afterglow SED measurements. For this experiment we set the FM90 coefficients $c_3$ and $c_4$ to their best-fit values of 0.0 and 0.43 respectively. Both these parameters also influence the UV slope of the extinction curve, and for now we therefore froze these in order to better illustrate the dependancy between $R_V$ and $c_2$. We explored the parameter range $2<R_V<6$, $0.4<c_2<3.6$, which covers the range in $R_V$ that has thus far been measured \citep{gcm+03,fm07}. The upper panel of Fig.~\ref{fig:rvc2cont_c3} shows the best fitting [$R_V$, $c_2$] pairs for the GRB SEDs together with their confidence contours. For comparison the average values of the [$R_V$, $c_2$] parameters found along SMC, LMC and Milky Way sightlines are also shown. With $c_3$ fixed at 0.0 we see that the mean [$R_v$, $c_2$] values for the SMC and the LMC lie within the 90\% confidence region, whereas the Milky Way is outside of the 99\% confidence region. This finding is consistent with the disparity between the mean GRB extinction curve and mean Milky May extinction curve, as previously illustrated in the Fig.~\ref{fig:meanextcurve} curves. Fig.~\ref{fig:rvc2cont_c3} confirms that for $c_3 = 0.0$ and $c_4 = 0.43$, we can rule out with 99\% confidence that the mean GRB UV extinction curve is as flat as the mean Milky Way curve.

To explore how much the presence of a \MWbump\ bump in the GRB extinction curves would affect our constraints on $R_V$, $c_2$, we repeated the previous experiment but with $c_3$ set to 1.0 (with $c_4$ fixed again to 0.43). The results are shown in the lower panel of Fig.~\ref{fig:rvc2cont_c3}. We see that, as before, the Milky Way [$R_V$, $c_2$] parameters lie outside the outer confidence contour, and now the LMC [$R_V$, $c_2$] parameters are also outside the 99\% confidence region. Therefore, even if the mean GRB extinction curve does have a prominent \MWbump\ bump, it is still incompatible at 99\% confidence with an extinction law that is as flat as the mean Milky Way extinction law, and it is also incompatible with being as flat as the LMC extinction curve.

Finally, we have also explored what constraints we can place on the $R_V-c_2$ parameter space with $c_4$ left as a free parameter, and we find that although with the extra free parameter the confidence regions widen, for both $c_3=0.0$ and $c_3=1.0$, the [$R_V$,$c_2$] Milky Way values are still excluded at 99\% confidence.

We now explore further the slope of the GRB extinction curves at rest-frame UV wavelengths for the sample of GRBs that were excluded from our final sample. We use the parameter $c_2$ (at a given $R_V=3.16$) as our primary measure of the UV-slope, where $\lambda$ is in microns, and for $\lambda<0.27\mu$m.

\subsubsection{Are we biased against flat extinction curves?}
\label{sssec:fltbias}
It is possible that our initial step of filtering out GRBs that apparently have a spectral break within their NIR to X-ray afterglow SEDs could preferentially reject sources with intrinsically flat dust extinction curves. This is because there is a chance that a flatter extinction law than that considered in our analysis may have been compensated by a best-fit optical spectral index that is flatter than the true value, as well as by a smaller total host reddening.

To investigate if our method is biased against flat dust extinction curves, we tested whether the GRBs with an apparent spectral break can alternatively be explained by unbroken power-laws with particularly flat dust extinction models. There were 24 GRBs in our sample with an apparent spectral break, and we re-fit these with a single power-law continuum but using the more flexible FM90 extinction curve prescription to model the host galaxy dust extinction. There are typically too few UV to NIR data points in these SEDs to constrain all the coefficients in the FM90 extinction model. The FM90 parameters were therefore fixed to the best-fit parameters from our simultaneous SED analysis (Table~\ref{tab:fitres}), with the exception of $c_1$, which was left free to vary, and which is tied to $c_2$.

The results of these new fits are summarised in Table~\ref{tab:refits}, where we also give the goodness of fit to the afterglow SEDs provided by a broken power law model with {\em normal} (i.e. SMC, LMC or Milky Way) host galaxy dust extinction. A broken power-law with a local dust extinction model has the same number of free parameters as a power-law with the FM90 extinction curve, and so it is acceptible to compare directly the minimum $\chi^2$ acheived by these two models. In only 5/24 cases (GRB~061121, GRB~061126, GRB~080413B, GRB~080721 and GRB~081203A) a power-law with the FM90 extinction curve provided a better fit to the data than a broken power-law with dust extinction modelled on the local Universe. These five GRBs are highlighted in bold in Table~\ref{tab:refits}. In the case of GRB~081203A, the best-fit $c_1$ parameter lies within the best-fit value to the mean Milky Way and mean LMC extinction curve (see Table~\ref{tab:fitres}). The improvement in the fit statistic relative to the SMC, LMC and Milky Way extinction law models is therefore likely to be due to the lack of a \MWbump\ feature in our `flat dust' extinction curve model. Furthermore, we note that at the epochs where we have extracted SEDs for GRB~061121, GRB~061126 and GRB~080721, the temporal decay rates measured in the X-ray band are different from those in the optical/NIR bands (GRB~061121; see fig.~10 of Page et al., 2007, GRB~061126; see fig.~4 \& 7 of Gomboc et al., 2008 and see fig.~6 \& 8 of Perley et al., 2008a, GRB~080721; see table~2 of Starling et al., 2009). The NIR through to X-ray afterglow emission for these three GRBs cannot therefore arise from the same emission component, and thus cannot lie on same NIR-X-ray spectral component.

This therefore just leaves GRB~080413B\footnotemark[6] as possibly having a host galaxy dust extinction curve that is flatter than those host extinction curves used in our SED analysis (section~\ref{sec:method}).
\footnotetext[6]{The X-ray and optical/NIR afterglow light curves of GRB~080413B follow a complex evolution that do not satisfy the closure relations and are best explained by a two-component jet \citep{fkg+11}. However, within this model, the emission at the epoch of our SED is likely to be dominated by a single jet component, in which our single component spectral model would be valid.}
This GRB has a best-fit $c_1$ coefficient $c_1=2.94^{+0.34}_{-0.39}$, which is much larger than the best-fit FM90 coefficients corresponding to the mean SMC ($c_1=-4.47\pm 0.19$), LMC ($c_1=-1.28\pm 0.34$) and Milky Way ($c_1=0.12\pm0.11$) values, and thus corresponds to a flatter extinction curve. This would suggest that only of order 2\% of GRBs that meet our initial parent sample selection criteria (49~GRBs; see section~\ref{ssec:sample}) have flat dust extinction curves.

A more conservative upper limit on the fraction of GRBs with a flat host extinction curve is given by considering the number of GRB afterglows acceptably fit by a flat FM90 model, independent of the goodness of fit of other dust models. For this, we only consider those GRBs with afterglows that decay at comparable rates in the X-ray through to optical/NIR bands, for the same reason as was discussed above for GRBs~061121, 061126 and 080721. I.e. it is only those GRB afterglows that have identical temporal behaviour in the X-ray and NIR/optical/UV bands that can have a single NIR to X-ray spectral component. We therefore compare the temporal decay rates measured in the X-ray band and the NIR-optical bands for each of the GRBs in Table~\ref{tab:refits} to identify what fraction of these GRBs cannot have a single power law continuum. We find that 14/24 of these GRBs have decay rates in the X-ray band that differ from their decay rates in the NIR/optical bands, indicating that they must have a continuum that is more complex than a single powerlaw. This leaves 10 GRBs that can potentially have a single NIR to X-ray spectral slope. A column is included in Table~\ref{tab:refits} to indicate whether the optical and X-ray decay rates of each GRB are consistent at the epoch of the SED. Of these, 7/10 were acceptably fitted by a power law and flat host galaxy extinction curve model at the 99\% confidence level, although a further single GRB (GRB~090618) was significantly better fitted with a spectral break. We therefore consider a maximum of 6/49 GRBs, or an upper limit of 12\% of GRBs in our sample to have flat host extinction curves.

\subsubsection{Redshift dependent extinction curves}
\label{sssec:zlmt}
Our initial sample of 49 GRBs had a redshift range $z=0.5-4.0$ with a median redshift of $\langle z\rangle=1.7$, and in our final sample of 17 GRBs, the redshift range was from 0.7 to 3.1 with a median of $\langle z\rangle=$1.4. The median redshift of our GRB sample is lower than the median of $\langle z\rangle\sim 1.94$ for the complete sample of \swift\ bursts\footnotemark[7].
\footnotetext[7]{http://www.raunvis.hi.is/~pja/GRBsample.html}
This difference is the result of our selection criteria, which favours bright afterglows, and thus (relatively) low-z GRBs. It is currently not clear whether the galaxy dust extinction curves change as a function of redshift for redshifts $z\lesssim 5$, and this bias in our sample should thus be kept in mind when applying the results presented in this paper to GRBs at redshifts $z\gg 2$.

\subsubsection{Host \av\ dependent extinction curves}
\label{sssec:fluxlmt}
A more obvious selection effect in our analysis is the bias against very dusty lines-of-sight. By requiring that the GRBs in our sample have both a spectroscopic redshift measurement and a UVOT $v$-band magnitude $v<19$ we remove GRBs from our sample with highly extinguished afterglows, both due to dusty regions within their host galaxy and/or high Galactic extinction along their line-of-sight. If the extinction law of GRB host galaxies varies with the host galaxy dust column density, then our analysis will clearly only be applicable to GRBs with a host galaxy visual extinction smaller than a certain threshold, above which the GRB afterglow would be too extinguished to enter our sample. There is empirical evidence that the prominence of the \MWbump\ extinction bump is related to the total visual extinction, \av, along the line-of-sight. The host extinction curves of GRB~070802, GRB~080605, GRB~080607 and GRB~080805 are the only ones with a spectroscopically detected \MWbump\ extinction bump, and the afterglows are also of the most heavily extinguished. This point is also discussed in \citet{zwf+11}.

Such a relation between the prominence of the \MWbump\ feature and visual extinction, \av, is not necessarily indicative of evolution in the dust extinction curve, and may be a result of there not being sufficient extinction in low dust environments for the \MWbump\ feature to be detected with any statistical significance. It is, nevertheless, possible that the results presented in this paper are only valid for the host galaxies of those GRBs that have a total visual extinction $A_V<1$. In view of this prospect, in the next section we investigate further the origin of differences in the host galaxy extinction curves of GRBs with only moderately extinguished afterglows, such as in our sample, and those that are highly extinguished (i.e. $A_V>1$).

\subsection{Variation in GRB host extinction laws}
\label{ssec:dstdst}
The improved positional accuracy of GRBs provided by the rapid response of \swift\ and (semi-)robotic ground-based telecopes, in particular those equipped with NIR instruments, has significantly improved our capabilities over the last half decade to study highly dust-extinguished GRBs and their host galaxies. Dedicated host galaxy follow-up programmes of highly dust-extinguished GRB candidates are revealing a sample of GRB host galaxies that are typically more massive, luminous, and chemically evolved than the typical host galaxies of relatively unextinguished GRBs (GRB~020127; Berger et al. 2007, GRB~030115; Levan et al. 2006, GRB~070802; Kr{\"u}hler et al. 2011, GRB~080325; Hashimoto et al. 2010, GRB~080607; Chen et al. 2011).

In the case of GRB~080607, \citet{psp+09} argued the majority of the afterglow dust extinction to come from dust within an intervening molecular cloud. However, it is notable that the characteristic properties of the host galaxy are also atypical when compared to optically-selected samples \citep{cpw+11}. The host galaxy of GRB~080607 had a stellar mass $M_*\sim 8\times 10^9~M_\odot$, which is almost an order of magnitude larger than the mean stellar mass of optically selected samples \citep[$\langle M_*\rangle\sim 10^9~M_\odot$;][]{sgb09,kgs+11}, and it was very red, with $R-K> 5$ \citep{cpw+11}. This is also the only GRB to have a robust detection of molecular hydrogen absorption in its afterglow spectrum \citep{psp+09}.

On the other hand, GRB~070306 and GRB~100621A were heavily extinguished GRBs ($A_V\sim 5.5$ and $A_V\sim 3.8$ respectively), but both had very blue host galaxies, with $R-K$ colours comparable to the host galaxies of relatively unextinguished GRBs \citep{jrw+08,kgs+11}, indicative of a very clumpy distribution of dust. In a sample of six highly extinguished GRBs, \citet{pcb+09} also found evidence for a patchy dust distribution. Furthermore, although GRB~070306 had one of the largest stellar masses measured for a long GRB host galaxy \citep[$2\times 10^{10}~M_\odot$;][]{kgs+11}, the host stellar mass for GRB~100621 ($10^{9}~M_\odot$) was comparable to that of optically bright afterglow host galaxies. The relation between afterglow dust extinction and host galaxy properties, including the host extinction curve, is therefore not clear, and further investigation of the host galaxies of heavily extinguished afterglows will be needed to address these issues.

\section{Conclusions}
\label{sec:conc}
Using a sample of 17 GRBs with good wavelength coverage and a single NIR through to X-ray spectral emission component, we have been able to constrain the host galaxy dust extinction properties as a function of wavelength for GRBs with total host galaxy visual extinction $A_V<1$. By studying the host galaxy dust extinction properties for a sample of GRBs rather than for a single, well-observed, case, the analysis presented in this paper allows us to investigate the average dust extinction properties in the host galaxies of GRBs, albeit only for those with a moderate host galaxy visual extinction ($A_V<1.0$). Using the \citet{fm90} analytic extinction curve model to fit our afterglow sample, we find the best-fit to have a slope blueward of $\sim 2000$\AA\ that is intermediate to that of the SMC and LMC mean extinction curves, and it has no \MWbump\ extinction bump. Within the uncertainties, the best-fit extinction law is consistent with both the SMC and LMC mean extinction curves. However, our data are inconsistent with having a host galaxy extinction curve as flat as the average Milky Way curve blueward of $\sim 2000$\AA.

Our best-fit FM90 parameterisation of the GRB host extinction curve provides an acceptable fit (at $3\sigma$ confidence) to a full sample of 49 afterglow SEDs that we initially analysed, providing typically a similar or improved fit to the afterglow SEDs compared to fits that use local dust extinction curve models. Of those GRBs with host galaxy dust extinction measured at 90\% confidence, around 10\% were, however, better fit by an extinction curve that contained the 2175\AA\ dust feature. Furthermore, up to 12\% of the 49 afterglow SEDs analysed could have a host galaxy extinction curve that is significantly flatter at rest-frame UV wavelengths than the mean Milky Way extinction curve. For this 12\% of GRBs, the improved fit provided by a flatter UV-extinction curve model increases the host galaxy total extinction relative to fits with local extinction curve models. This is compatible with the trend increasingly observed between significantly extinguished GRBs ($A_v\gg 1$) and flat host galaxy extinction curves.

The relatively UV-steep host galaxy extinction curve for our sample of modestly extinguished GRBs, and the much flatter extinction curves with more pronounced 2175\AA\ features of host galaxies of more extinguished GRBs, is adding to the empirical evidence that the extinction curve of GRB hosts vary as a function of host dust abundance and/or some other galaxy property. In view of this, a future test would be to study the average GRB host extinction curve properties for a sample of GRB afterglows with little  (e.g. $A_V<0.3$) and high ($A_V>1.0$) host galaxy extinction. Such analysis would require a sample of bright, relatively unextinguished GRBs in the former case, and a sample of extinguished GRBs with good broadband rest-frame NIR through to X-ray coverage in the latter, and preferably with no spectral break within the observed bandpass. The current size of such a sample remains relatively small. However, examples of highly extinguished cases such as GRB~080607 show that such GRB samples are attainable with time, as long as rapid-response observing programmes that cover the optical through to NIR afterglow remain available.

\section*{ACKNOWLEDGEMENTS}
We thank the referee for the helpful comments that have improved this paper. P.S. acknowledges support through project SA 2001/2-1 of the DFG. T.K. acknowledges support by the Deutsche Forschungsgemeinschaft (DFG) cluster of excellence ÕOrigin and Structure of the UniverseÕÊas well as support by the European Commission under the Marie Curie Intra-European Fellowship Programme. The Dark Cosmology Centre is funded by the Danish National Research Foundation. S.R.O. and M.J.P. acknowledge the support of UKSA. This research has made use of data obtained from the High Energy Astrophysics Science Archive Research Center (HEASARC), the UK Swift Science Data Centre at the University of Leicester and the Leicester Data base and Archive Service (LEDAS), provided by NASAs Goddard Space Flight Center and the Department of Physics and Astronomy, Leicester University, UK, respectively.

\clearpage
\appendix
\section{SED fits}
\begin{figure*}
\includegraphics[width=1.0\textwidth,angle=-90]{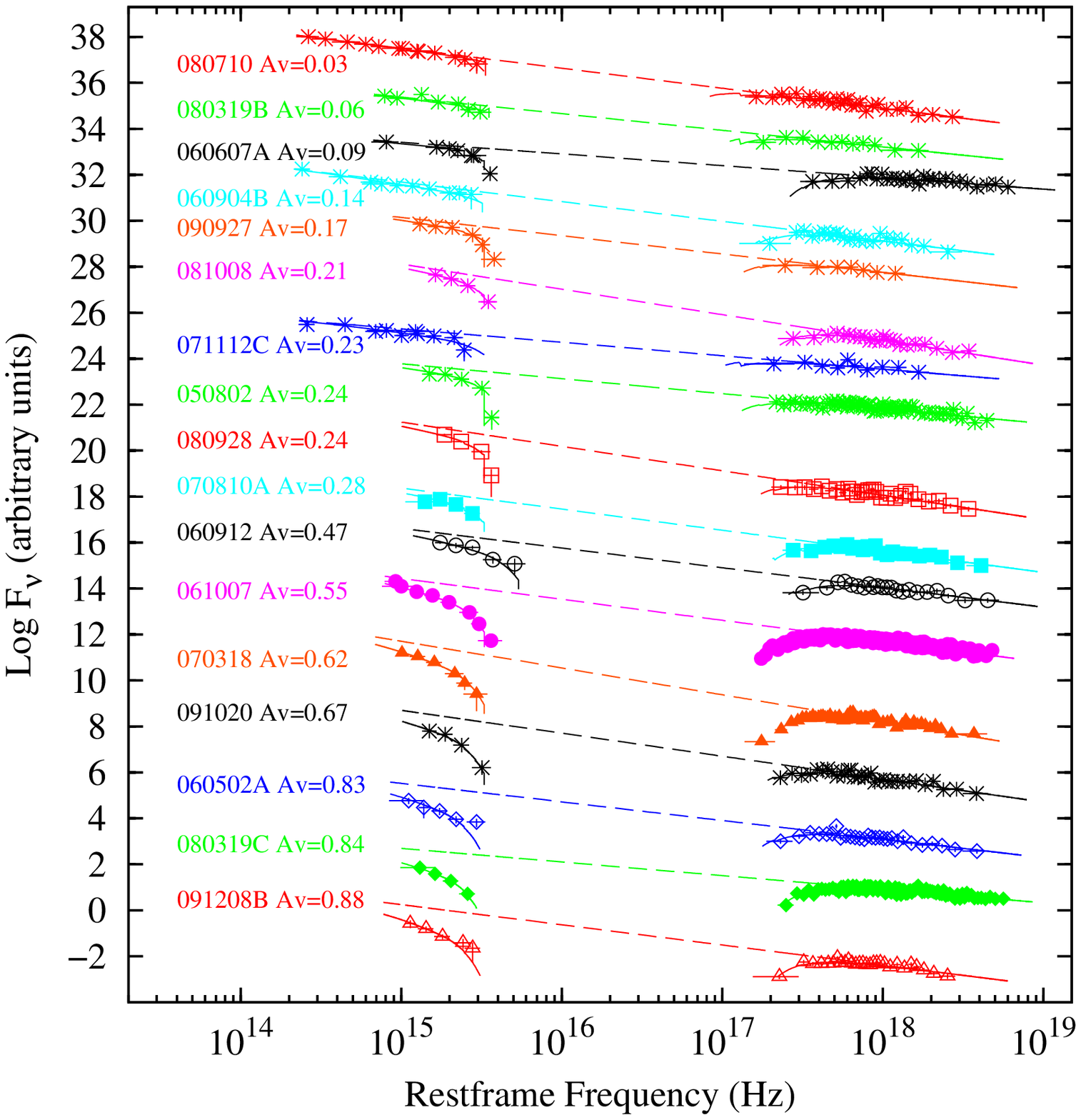}
\caption{NIR through to X-ray SEDs of the 17 GRBs in our final sample with the best-fit FM90 model overplotted. Frequency is plotted in the rest-frame and flux density is in arbitrary units for clarity. The SEDs have been corrected for Galactic reddening and soft X-ray absorption, and both the host galaxy absorbed/extinguished (solid) and intrinsic (dashed) best-fit afterglow SED are indicated. The afterglow SEDs are ordered by afterglow host galaxy extinction, with the most extinguished at the bottom, and the least at the top.}
\label{fig:fnlSEDs}
\end{figure*}

\begin{figure*}
\includegraphics[width=0.9\textwidth,angle=-90]{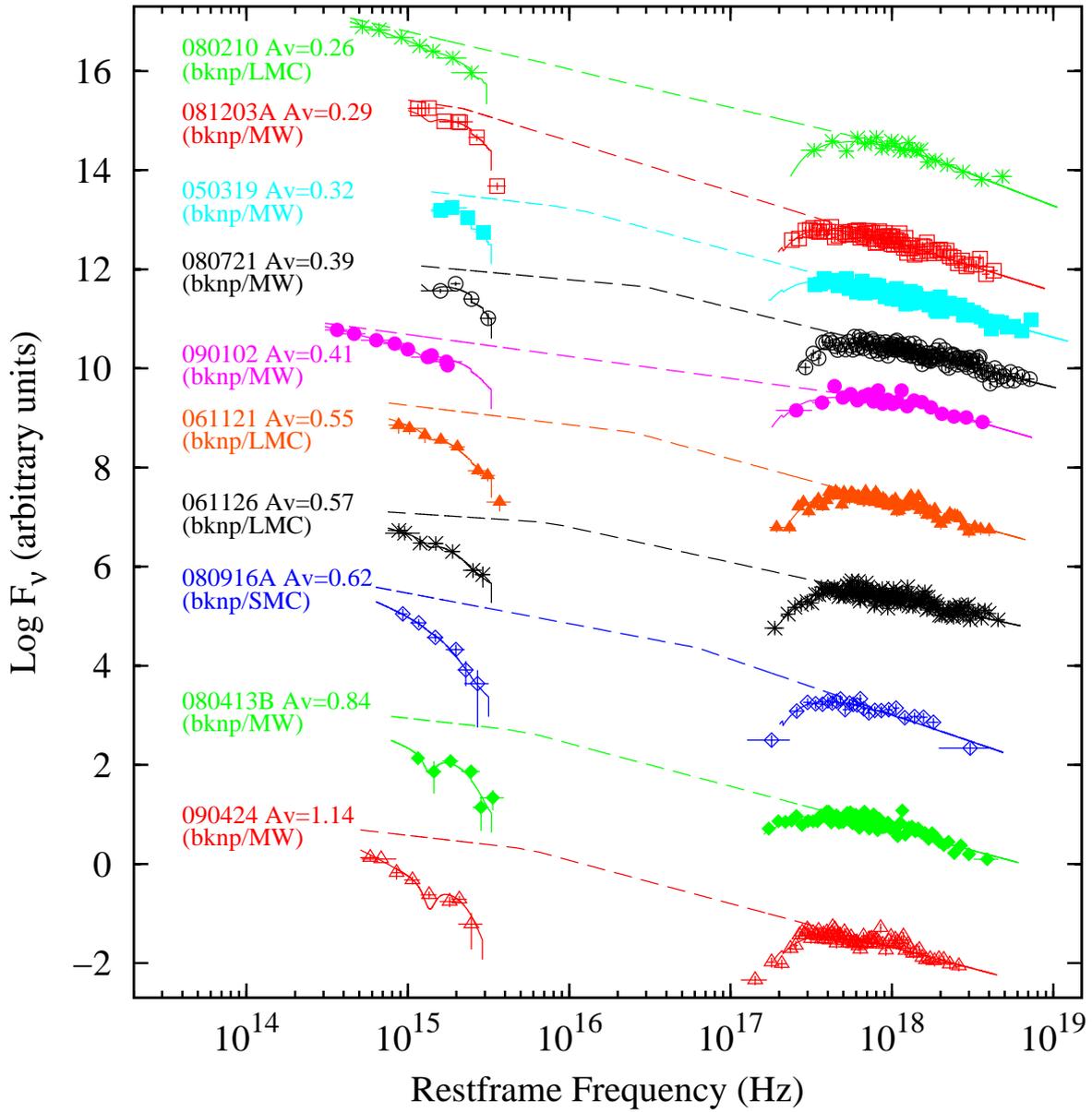}
\caption{NIR through to X-ray SEDs of the 32 GRBs analysed that did not enter our final sample. Frequency is plotted in the rest-frame and flux density is in arbitrary units for clarity. The SEDs have been corrected for Galactic reddening and soft X-ray absorption, and both the host galaxy absorbed/extinguished (solid) and intrinsic (dashed) best-fit afterglow SED are indicated. The afterglow SEDs are ordered by afterglow host galaxy extinction, with the most extinguished at the bottom, and the least at the top.}
\label{fig:SEDs}
\end{figure*}

\begin{figure*}
\setcounter{figure}{1}
\includegraphics[width=0.9\textwidth,angle=-90]{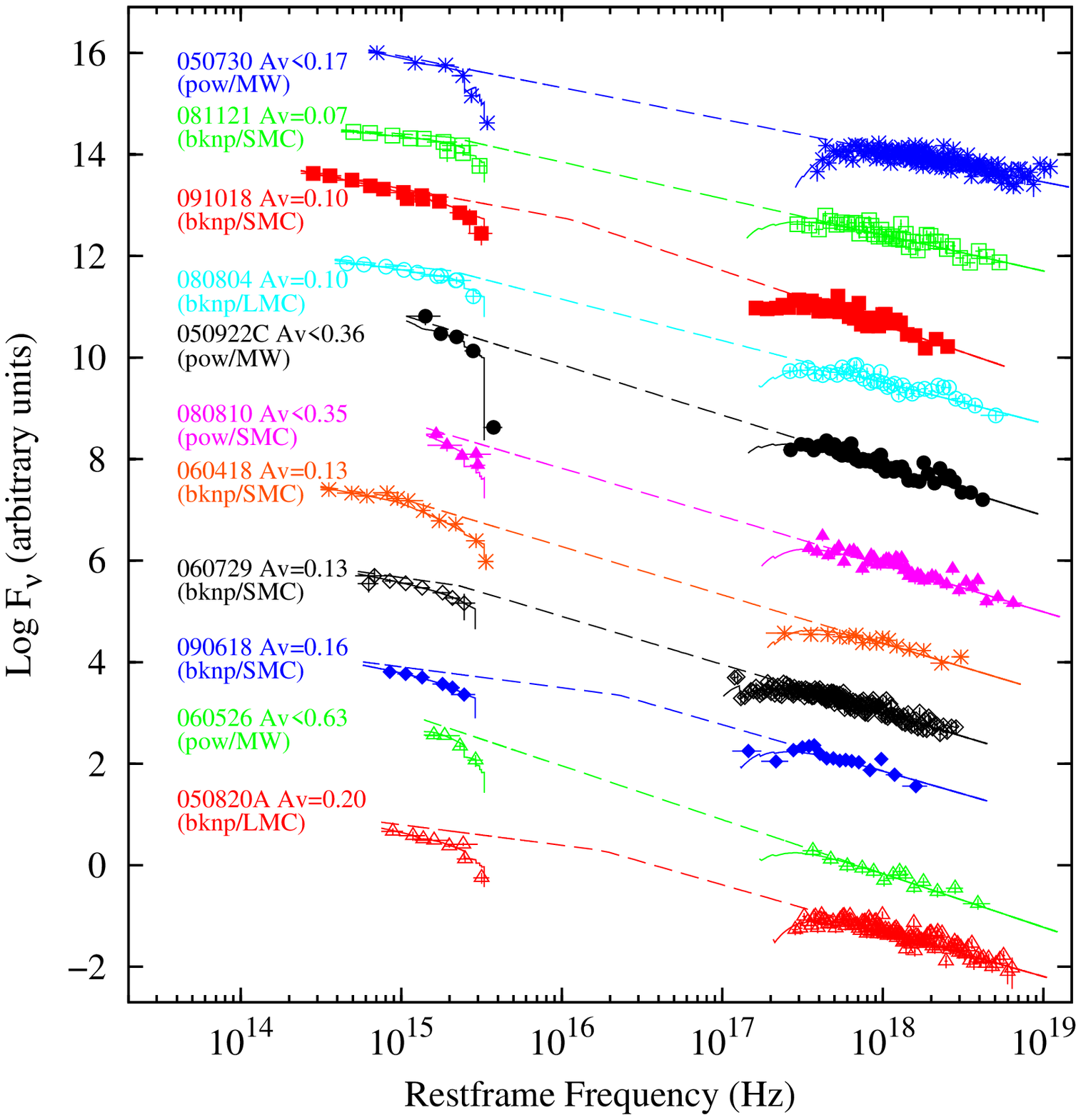}
\caption{(continued)}
\end{figure*}

\begin{figure*}
\setcounter{figure}{1}
\includegraphics[width=0.9\textwidth,angle=-90]{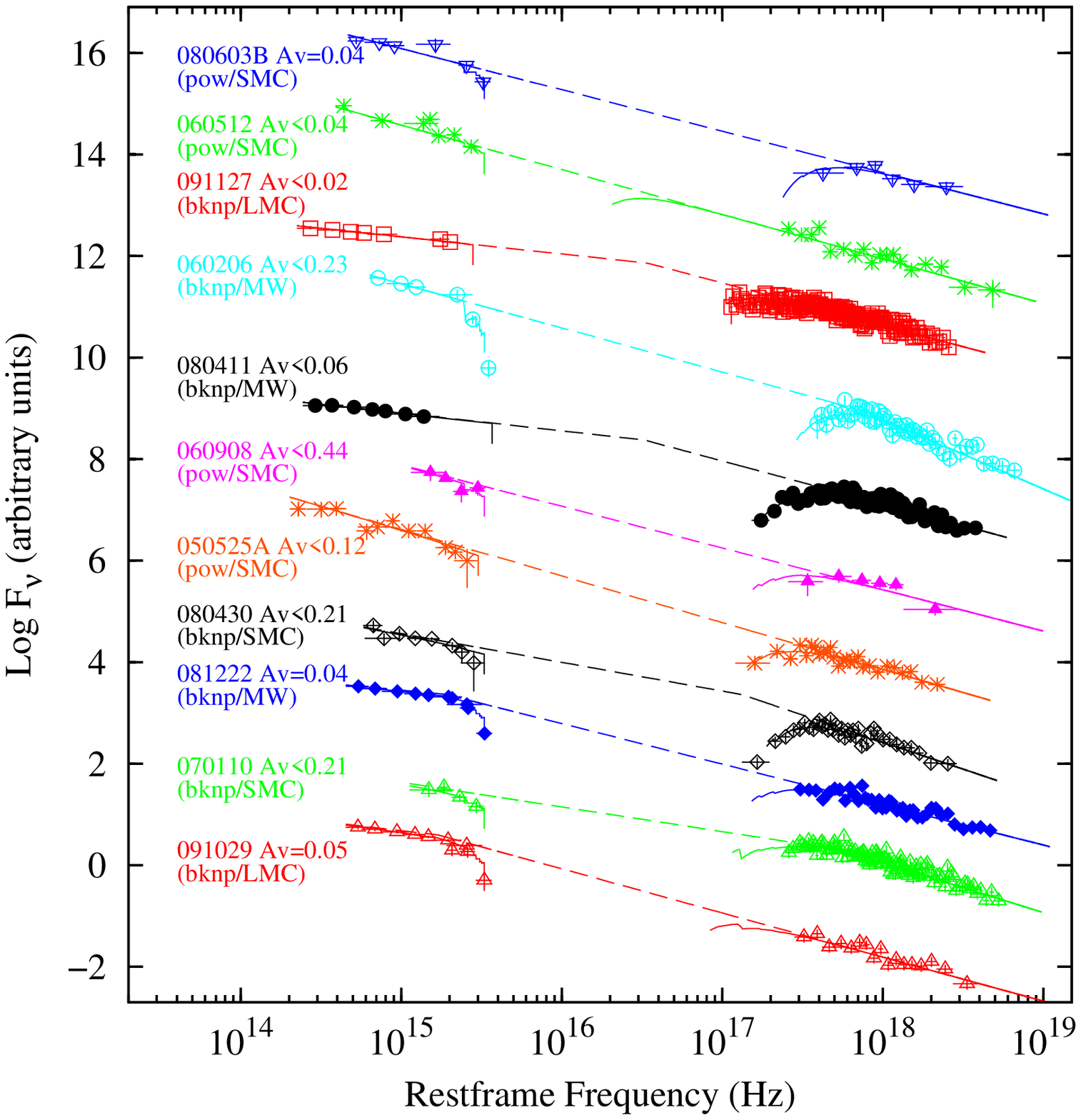}
\caption{(continued)}
\end{figure*}

\end{document}